\documentclass[runningheads]{llncs}

\usepackage[utf8]{inputenc}
\usepackage{main}

\begin{document}

\bibliographystyle{splncs04}

\title{SMT Sampling via Model-Guided~Approximation\thanks{This work is supported by the Israeli Science Foundation Grant No. 243/19 and the Binational Science Foundation (NSF-BSF) Grant No. 2018675.}}

% DOUBLE BLIND
\author{Matan~I.~Peled\inst{1}\orcidID{0000-0001-8899-3235} \and Bat-Chen~Rothenberg\inst{1}\orcidID{0000-0001-8345-7273} \and Shachar~Itzhaky\inst{1}\orcidID{0000-0002-7276-7644}} % FIXME
\institute{Technion---Israel Institute of Technology, Haifa, Israel \email{\{mip,shachari,batg\}@cs.technion.ac.il}} % FIXME

%\author{\IEEEauthorblockN{Matan I. Peled, Bat-Chen Rothenberg, and Shachar Itzhaky\\}
%\IEEEauthorblockA{Technion - Israel Institute of Technology\\
%\{mip,batg,shachari\}@cs.technion.ac.il}}
% \and
% \IEEEauthorblockN{Bat-Chen Rothenberg}
% \IEEEauthorblockA{Technion - Israel Institute of Technology\\
% Email: batg@cs.technion.ac.il}
% \and
% \IEEEauthorblockN{Shachar Itzhaky}
% \IEEEauthorblockA{Technion - Israel Institute of Technology\\
% Email: shachari@cs.technion.ac.il}}

\maketitle

\begin{abstract}
We investigate the domain of satisfiable formulas in
satisfiability modulo theories (SMT), in particular,
automatic generation of a multitude of satisfying assignments
to such formulas.
Despite the long and successful history of SMT in model
checking and formal verification, this aspect is relatively
under-explored.
Prior work exists for generating such assignments, or \emph{samples},
for Boolean formulas and for quantifier-free first-order formulas involving bit-vectors, arrays, and uninterpreted functions (QF\_AUFBV).
We propose a new approach that is suitable for a theory $T$ of integer
arithmetic and to $T$ with arrays and uninterpreted functions.
The approach involves reducing the general sampling problem to
a simpler instance of sampling from a set of independent
intervals, which can be done efficiently.
Such reduction is carried out by expanding a single model---a 
\emph{seed}---using top-down propagation of constraints along
the original first-order formula.

 % FIXME

\end{abstract}

\section{Introduction}\label{sec:intro}
% Outline:
%
% SMT formulas are often used for the analysis of hardware and software systems:
% examples a-c For all the above applications, obtaining multiple solutions is a
% desirable thing: explain in this paper, we focus on sampling: obtaining
% multiple diverse solutions for SMT While sampling has been widely investigated
% in the sat domain, not so in smt.  Sat is not enough because: a. it is not
% always possible to reduce to sat, b. even when possible, sampling on the smt
% level is more efficient.  Also, sampling one-by-one is inefficient.  So, we
% need efficient samplers on the smt level that generalize for many theories
% (especially over infinte domains).
%
% SMT formulas are often used for the analysis of hardware and software systems:
% examples: CRV, software testing, software verification
Satisfiability Modulo Theories (SMT) formulas are the centerpiece of many
modern-day algorithms for the testing and verification of hardware and software
systems. In constrained-random verification (CRV)~\cite{CRV} --- one of the
most popular methods for hardware testing in the industry --- the functional
model and verification scenarios of a hardware design are translated into an SMT
formula.
% in the theory of arrays, uninterpreted functions and bit-vectors.  Software
% testing methods - whitebox fuzzing / dynamic test generation
In software verification~\cite{cbmc,ultimate,seahorn}, SMT formulas are used to
express safety requirements extracted from the code.
% The solution to the satisfiability problem for that formula then indicates
% whether a safety violation can, in fact, occur, and a model\BC{terminology:
% satisfying assignment/structure? solution?}, if found, represents program
% inputs that trigger it.

% For all the above applications, obtaining multiple solutions is a desirable
% thing: explain
The problem of SMT solving has been widely investigated, and many solvers are
available~\cite{DeMoura:2008,Barrett:2011,Cimatti_Griggio_Schaafsma_Sebastiani_2013,Dutertre:cav2014}.
These tools can determine satisfiability and return a (single) model if it
exists. However, there are use cases in which multiple, sometimes multitudes of,
such models are needed. For example, in CRV, solutions of the formula represent
stimuli for the design under test, and multiple and diverse solutions increase
the likelihood of discovering
bugs~\cite{CRV}. %TODO: expand to general sampling (also software)
In the context of software verification, a solution of the formula often
represents an input causing a safety violation, along with the buggy execution
of the program for that input~\cite{cbmc,ultimate}. Obtaining multiple such
solutions provides additional insight into the bug in question, which may help
guide the debugging process.
% Furthermore, if the bug is repaired using
% automated test-based techniques such
% as~\cite{Le_Goues_Nguyen_Forrest_Weimer_2012,Mechtaev_Yi_Roychoudhury_2016},
% additional test cases become crucial for their
% performance~\cite{KongReport,Yang_Zhikhartsev_Liu_Tan_2017}\BCFV{Check that this
%   citation indeed supports this claim...}.

This paper is concerned with the \textit{SMT sampling} problem, \ie efficiently
generating multiple random solutions for an SMT formula with good coverage of
the solution space.  The na\"ive solution of enumerating models using a solver
often becomes too expensive in practice, and most solvers tend to return similar
models in successive invocations.  Notably, while the problem of \emph{SAT}
sampling has been successfully established with techniques such as Markov~Chain
Monte-Carlo~(MCMC)~\cite{kitchen2007stimulus,kitchen2010markov} and universal
hashing~\cite{meel2014sampling,meel2016constrained,ermon2013embed}, the history
of the same problem applied to SMT formulas is
shorter~\cite{Dutra:2018,Dutra:2019}. When considering propositional formulas or
formulas over bit-vector theories, a reduction from SMT sampling to SAT sampling
is possible. However, it was shown in~\cite{Dutra:2018} that such an approach is
significantly less efficient than sampling at the SMT level directly.
Furthermore, when considering formulas over infinite domains, such as the
integers or the reals, a reduction to SAT is not even an option.
% 
%Thus, there is an urgent need for efficient SMT sampling techniques that
%generalize to a variety of theories, and in particular to theories where the
%domain is infinite.

This paper presents a novel algorithm for sampling SMT formulas in the theory of
linear integer arithmetic enhanced with (possibly non-linear) multiplication,
denoted $\Tmia$.  We also present an extension to integer arrays and
uninterpreted functions.  The extended theory is denoted $\Tamia$.  Our approach
is \textit{epoch-based}~\cite{Dutra:2018}: it operates in a series of rounds,
called \emph{epochs}, where in each epoch a \textit{seed} $m$, which is a random
model of the formula, is generated using an off-the-shelf solver.  This model is
then extended to a large set of models of similar nature, in a cost-effective
manner.

The novelty of our approach lies in the algorithm for extending the model $m$
and in the representation of the set of models returned in each epoch.  This
algorithm relies on the novel notion of \textit{model-guided approximation}
(MGA), which we introduce.  MGA uses a model $m$ of a formula $\varphi$ to
derive a simpler formula, $\varphi'$, s.t. $m\models \varphi'$ and $\varphi'$
\emph{underapproximates} \(\varphi\).
% In addition, MGA is parameterized in two theories, $T$ and $T'$, where $T$ is
% the theory $\varphi$ is taken from, and $\varphi'$ must be taken from $T'$.
% In our use of MGA for sampling, $m$ is the seed, $\varphi$ is the input
% formula, and $T'$ is chosen to be the theory of intervals, as formulas in this
% domain can be sampled efficiently and uniformly.  Thus, the resultant formula
% $\varphi'$ can be sampled efficiently, and the underapproximation property
% ensures that every collected solution is also a solution of \(\varphi\).  The
% key step is generating a formula \({\varphi}'\) from \(\varphi\), such that
% \(\varphi'\) is a conjunction of interval constraints over each variable,
% $m\models \varphi'$ and $\varphi'$ \emph{underapproximates} \(\varphi\).
% Intuitively, if you think of \(\varphi\) as the set of all its solutions and
% of $m$ as an element in this set, then \({\varphi}'\) is a subset of
% \(\varphi\) that contains $m$.
The rationale is that using $m$ during the underapproximation process can help
guide it towards solutions that are similar to $m$, and avoid parts of the
search space which are devoid of solutions.
%As a simple example, consider a formula of the form $\varphi=c_1\vee c_2$, where $c_1$ and $c_2$ are arbitrary
%complex constraints. Without any additional knowledge, one might choose to
%underapproximate $\varphi$ using either $c_1$ or $c_2$, either of which could be
%unsatisfiable. A model that satisfies $c_1$, on the other hand, gives us the
%confidence that $c_1$ is satisfiable, which increases our hope to find solutions
%focusing on $c_1$ as an approximation.

Our sampling algorithm uses MGA in every epoch to convert a $\Tmia$-formula
$\varphi$ into a fomrula $\varphi'$ in a theory of intervals, $\Tic$. In $\Tic$,
formulae are restricted to a conjunction of constraints of the form $x\ge c$ or
$x\le c$, where $x$ is a variable and $c$ is a constant. The underapproximation
is done using a rule-based approach, which propagates constraints in a top-down
fashion along the abstract syntax tree (AST) of the formula. Obtaining a set of
concrete solutions from $\varphi'$ is straightforward: by repeatedly sampling
all variables from within their boundaries. Note that the underapproximation
property of MGA ensures that every point within these intervals is necessarily a
model of $\varphi$.

The ability of our algorithm to represent the set of solutions at each epoch
symbolically in the form of an interval formula $\varphi'$ has several
advantages. First, it allows blocking all previously seen solutions by
conjoining $\varphi$ with $\neg\varphi'$. Such blocking is often feasible since
the size of $\varphi'$ is proportional to the number of variables (rather than
the size of the solution set, which can be large or even infinite).  On top of
that, the interval formula itself can be returned instead of a concrete set of
solutions, which can give added value to the user.  For example, in the scenario
where solutions of the formula represent inputs that cause a bug in a program,
knowing that every $x$ in a certain range causes the bug can be helpful for
debugging.

Our algorithm does not aim to provide formal guarantees regarding uniform
sampling nor coverage. This is in contrast to some prior work on SAT
sampling~\cite{Chakraborty_Meel_Vardi_2013,meel2016constrained,ermon2013embed},
but similar to prior work on SMT sampling~\cite{Dutra:2018,Dutra:2019}.  We
believe that the ability to adjust the sampling method towards an
application-specific goal, as opposed to a universal metric, is more important
than approximating a uniform distribution, since the purpose of sampling varies
by use case.  Our algorithm therefore allows to control the sampling heuristic
via two parameters: the choice of the initial seed, and the sampling of each
interval formula. Here, too, the use of intervals have the advantage of being a
convenient representation, in which it is easy to apply diverse sampling
heuristics such as uniform or even exhaustive sampling (if the space is finite).
\BC{examples: path coverage, wire-coverage, shortest-path-bug}

% Putting it all together, we obtain an epoch-based sampling algorithm for
% formulas in the theory of linear arithmetic enhanced with multiplication,
% which is based on the sampling of formulas in the interval domain.
We have implemented our algorithm in an open-source tool,
\MeGASampler\footnote{Available at: \url{https://github.com/chaosite/MeGASampler}}.
In order to compare it with
state-of-the-art sampler \SMTSampler~\cite{Dutra:2018}, we have also implemented
an adaptation of their algorithm (originally designed for the theory of
bit-vectors with arrays and uninterpreted functions) to integers. We provide an
experimental evaluation of their method and ours on a large set of benchmarks
from SMT-LIB.
% Our results show that \MEGASampler is able to produce tens of thousands of
% samples for a large variety of complex formulas. \BC{taken from applications
% such as...}  Furthermore, the coverage obtained by \MEGASampler, according to
% a previously established metric~\cite{Dutra:2019}, is very high, and is higher
% than that of \SMTSampler for most cases.  This indicates that \MEGASampler is
% competitive with state-of-the-art in terms of both the number of solutions and
% their quality.\BC{change based on the results}
Our results show that \MeGASampler significantly improves state-of-the-art in
terms of both the number of solutions and their quality, as measured
in~\cite{Dutra:2018}. \BC{add explanation of metric and motivation?}

% 3. Benefits of the technique are more solutions to the original formula,
% especially with integer arithmetic for which the SMTSampler approach is not
% effective.
%
% Benefits of the technique are more solutions to the original formula,
% especially with integer arithmetic for which the \SMTSampler approach is not
% effective.  \MIP{Write more about benefits and evaluation}
%
% My draft for benefits:
% What they all have in common: The by-product of our algorithm, i.e., the
% intervals (formula in $T'$) for each epoch, can be of use.
% a. As many solutions per epoch as you'd like.  Moreover, we can open the
% black-box of sampling: if a particular interval is of interest to you you can
% decide to get more samples from it, and vice versa (if it is not interesting).
% More control over the quantity-quality trade-off.
% b. The ability to obtain samples in groups using formulas in a theory of
% interest (intervals in our case). A versatile notion of a sample.
% c. Efficient blocking (only if we test it in experiments)
% d. Explainability: an interval is a descriptive human-readble way to "explain"
% why a model satisfies a formula.  For example, if the model describes a bug in
% the system then the interval tells you which variables are irrelevant (those
% who are unlimited) and which values are problematic for the relevant ones.

To sum up, our main contributions are:
\begin{enumerate}[leftmargin=0.45cm,topsep=0pt]
  \item Define the problem of Model-Guided Approximation (MGA) for pairs of
        first-order theories $T$ and $T'$.
  \item Present an algorithm for computing MGA of an integer
        theory onto the theory of intervals, with
        support for arrays and uninterpreted functions.
        On top of this, we implement an epoch-based procedure for sampling formulas in the source theory.
        % \item Present two variants of the algorithm for integer
        % theories with and without arrays and uninterpreted functions.
        % a rule-based method for solving the model-guided approximation
        % problem for the theory of linear arithmetic enhanced with multiplication
        % and the theory of intervals. This method is then used to instantiate the
        % general sampling framework, forming an algorithm for sampling formulas
        % in the former theory via sampling of formulas in the latter.
  \item Implement the algorithm in an open-source sampling tool, \MEGASampler,
        and evaluate it against an integer-based variant of the state-of-the-art
        sampling tool, \SMTSampler, on a large set of SMT-LIB benchmarks.
\end{enumerate}

% 3c. Example formula
\subsection{Motivating Example}
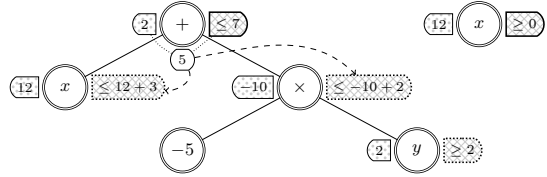
\begin{figure}[t]
  \centering
  \resizebox{0.6\textwidth}{!}{
  \begin{tikzpicture}[
    label distance = 0.75mm,
    sibling distance = 13em,
    level distance = 3.5em,
    every label/.style = {rounded rectangle,rounded rectangle arc length=120,
      font=\scriptsize, pattern color=black!20},
    model/.style = {draw,rounded rectangle right arc=none,pattern=crosshatch dots},
    constraint/.style = {draw,rounded rectangle left arc=none,pattern=crosshatch},
    syn/.style = {draw,double,circle,minimum size=2.4em},
    slack/.style = {draw,label distance=0.5mm}
    ]
    \node [syn,label={[model,name=two]left:{\(2\)}},
               label={[constraint,thick,name=seven]right:{\(\le 7\)}},
               label={[slack,name=slack]below:{\(5\)}}] (root) {\(+\)}
      child {node [syn,
             label={[model]left:{\(12\)}},
             label={[constraint,densely dotted,thick,name=const1]right:{\(\le 12+3\)}}] {\(x\)}}
      child {node [syn,
                   label={[model]left:{\(-10\)}},
                   label={[constraint,densely dotted,thick,name=const2]right:{\(\le -10+2\)}}] {\(\times\)}
        child {node [syn]  {\(-5\)}}
        child {node [syn,
                     label={[model]left:{\(2\)}},
                     label={[constraint,densely dotted,thick]right:{\(\ge 2\)}}] {\(y\)}}
    };
    \draw[densely dotted] (two) to[out=-60,in=150] (slack);
    \draw[densely dotted] (seven) to (slack);
    \draw[dashed,->] (slack) to[out=-60,in=-5] (const1);
    \draw[dashed,->] (slack) to[out=10] (const2.135);
    \node [syn,right=14em of root,label={[model]left:{\(12\)}},
    label={[constraint,thick]right:{\(\ge 0\)}}] {\(x\)};
  \end{tikzpicture}
  }
  \caption{Annotated syntax tree of \((x-5y\le 7)\land (x \ge 0)\) with model \( \{x \mapsto 12, y \mapsto 2\} \) }\label{fig:intro-example}
\end{figure}

As an introductory example, consider the integer formula:
\[ \varphi\colon (x-5y\le 7)\land (x \ge 0) \]

\MeGASampler begins its first epoch by consulting an off-the-shelf SMT~solver;
let us assume that the solver returned that
\( m = \{x \mapsto 12, y \mapsto 2\} \) is a valid solution to \(\varphi\). To
get more solutions from this seed solution, we can under-approximate \(\varphi\)
with an interval formula \(\varphi'\), which is easier to sample. In
\cref{fig:intro-example}, we see how this is done.

\Cref{fig:intro-example} shows an annotated syntax tree of \(\varphi\): to the left
side of each node is its value per the model $m$; to the right, we show the
bound for this term.  Solid outlines indicate that the inequalities are taken
directly from the formula, as is the case for the root nodes. Bounds with dotted
outlines are an inferred under-approximation.
%%%
Bounds are propagated in a top-down manner until the leaf nodes (integer
variables) are reached.  For example, the value ``$5$'' written below the
addition node represents the amount of ``slack'' to be distributed among child
nodes, as illustrated by the dashed arrows. By gathering the constraints on the
leaves (shown with thick borders) we obtain
\(\varphi'=x\le 15 \land y\ge 2 \land x \ge 0\), which represents the set of
intervals \(\{x\in\left[0,15\right],y\in\left[2,\infty\right]\}\).  Note that,
any solution that satisfies \(\varphi'\) also satisfies \(\varphi\).  Sampling
these intervals is then straight-forward: we can choose any value for \(x\) and
\(y\) in the intervals and it will be a valid solution to \(\varphi\). This
example is discussed in further detail in \cref{subsec:rules}.

%%% Local Variables:
%%% mode: latex
%%% coding: utf-8
%%% TeX-master: "main"
%%% TeX-PDF-mode: t
%%% TeX-engine: default
%%% End:
\section{Preliminaries}\label{sec:prelim}

% A first-order language consists of variables, logical symbols (such as $\land$, $\lnot$), and non-logical symbols (such as $+$, $-$, $<$), each with an \emph{arity} and a \emph{sort}.
% The set of non-logical symbols with their sorts comprises a \emph{signature}.
% The variables and symbols give rise to a set of formulas, constructed using a standard tree-like grammar. 

% A first-order theory~$T$ consists of a set of variables, a signature $\Sigma(T)$, a grammar $G(T)$,
% % a function mapping a fixed domain to each variable symbol,
% and a fixed interpretation for the symbols of the signature.\footnote{Some textbook theories include uninterpreted symbols; we exclude them from our presentation for simplicity.}

A first-order theory~$T$ consists of: 
a set of variables, each mapped to a concrete domain; 
a set of logical symbols (such as $\land$, $\lnot$); 
a set of non-logical symbols (such as $+$, $-$, $<$), each with an \emph{arity} and a \emph{sort}, called the \textit{signature} of $T$, denoted $\Sigma(T)$; 
a grammar $G(T)$;
a fixed interpretation for the symbols of the signature.
In addition, a theory may admit \emph{uninterpreted} function and predicate symbols, which appear in $G(T)$ but do not have fixed interpretations. These symbols are not considered to be part of $\Sigma(T)$.

The set of logical symbols is a subset of $\{\land,\lor,\lnot\}$ and their interpretation is fixed across all theories to be the standard one.
Note that, the interpretation of the non-logical symbols of the signature is also fixed, but depends on the theory. 

A \emph{structure}~$m$ consists of a domain, an interpretation of the symbols in the signature and an assignment of domain elements to all 
variables of the formula.

In the following sections, we will use $\Tlia$---the
standard theory of linear integer arithmetic; $\Tmia$,
which extends $\Tlia$ by allowing variable
multiplication (but not division);
and $\Tic$, which restricts $\Tlia$ to conjunctions of inequalities of the form $v\leq \IntConst$, $v \geq \IntConst$ (and strong variants with $<$, $>$).
We also consider extended theories that additionally admit uninterpreted functions and arrays, $\Tamia$ and $\Taic$.

%%% Local Variables:
%%% mode: latex
%%% coding: utf-8
%%% TeX-master: "main"
%%% TeX-PDF-mode: t
%%% TeX-engine: default
%%% End:

\section{Model-Guided Approximation}\label{sec:mbua}
In this section, we define the problem of \emph{model-guided approximation} (MGA).
% s
Consider a formula~\(\varphi\) in a theory~\(T\), and a model~\(m\) of
\(\varphi\). \(\varphi\)~can be seen as a representation of the set of
structures satisfying it,
\(\Models(\varphi)\triangleq\left\{m\mid m \models\varphi\right\}\).
Model-guided approximation aims to find a
subset~\(M\subseteq\Models(\varphi)\), which contains \(m\), and can be
represented using a formula~\(\varphi'\) that is better than \(\varphi\) for
some criteria. For example, for the purpose of sampling discussed in this paper,
\(\varphi'\) should be easier to sample than \(\varphi\).  Alternatively, one
can think of other goals, such as making~\(\varphi'\) human-readable or
easier to solve than~\(\varphi\).

To make sure \({\varphi}'\) is more suitable than \(\varphi\) in the criteria,
\({\varphi}'\) is limited to a theory~\(T'\). The theory~\(T'\) can be chosen
a-priori based on the criteria, and can be considered to be a part of the
problem. Intuitively, \(T'\) restricts \(T\) by adding syntactic limitations to
the way formulas are built, but does not change the semantics of
operations. Formally, the restriction relation is defined as follows:

\begin{definition}[Theory restriction]\label{def:theory-restriction}
A theory $T'$ \emph{restricts} $T$, denoted \(T'\restricts T\), if:
\begin{itemize}
  \item The language of the grammar of $T'$ is a subset of the language of the
        grammar of $T$. That is, every $T'$-formula is also a $T$-formula. Note
        that this requires, in particular, that $\Sigma(T')\subseteq\Sigma(T)$.
  \item The set of variables and uninterpreted function symbols (if any) as well as their mapped domains are identical in $T$ and $T'$.
  \item Every symbol $\sigma\in\Sigma(T')$ (which also belongs to $\Sigma(T)$)
        has the same fixed interpretation in both theories. 
\end{itemize}
\end{definition}

\begin{example}
  The theory $\Tic$, described in \cref{sec:prelim}, restricts both theories
  $\Tlia$ and $\Tmia$, which are also described there.  The common interpreted
  symbols, $<$, $\leq$, $>$, $\geq$, and all integer concepts, have the same
  interpretations, and any $\Tic$-formula is also a $\Tlia$-formula, as well as
  a $\Tmia$-formula.
\end{example}

\begin{definition}[Entailment]
  Semantic entailment is defined as a binary relation $\Rightarrow$ over formulas such
  that $\varphi\Rightarrow\psi$ iff for every structure $m$, if $m\models\varphi$ then $m\models\psi$.
\end{definition}

That is all well and good when referring to formulas of the same theory.  In the
presence of multiple theories, the situation is a bit more subtle.  Let
$\varphi$ be a $T'$-formula, $\psi$ a $T$-formula, and $T'\preceq T$. For a
$T'$-structure $m$, we denote $m^T$ its \emph{extension} to T, naturally
obtained by filling in any symbols that are not assigned in $m$ with their fixed
interpretations according to $T$ (all uninterpreted symbols are already
assigned interpretations in $m$, as follows from the previous definitions).
This allows for a slightly adjusted definition of $\Rightarrow$, namely:

\begin{definition}[$\preceq$-Entailment]
For $\varphi$,$\psi$ as above,
$\varphi\Rightarrow\psi$ iff for every $T'$-structure $m$,
if $m\models\varphi$ then $m^T\models\psi$.
\end{definition}

% \textbf{TODO}: explain the intuition behind the definition: syntax and
% semantics. every formula in the small is also a formula in the large.  Note that
% formulas in $T'$ are syntactically limited compared to $T$, but the semantics
% of their conjoined parts remain the remain the same.  The first three conditions
% express that formulas in $T'$ may be syntactically limited compared to~$T$:
% $T'$ allows to use only a subset of the symbols of $T$ (both logical and
% non-logical).  In particular, every formula in $T'$ is also in $T$, but not the
% other way around.\MIP{This can be done using a running example?}

To formally define model-guided approximation, we begin with the notion of
\emph{\(m\)-approximation}:
\begin{definition}\label{def:m-approximation}
  Let \(\varphi\) be a formula in a theory~\(T\) and \(m\) be a model of
  \(\varphi\). A formula~\({\varphi}'\) is called an \emph{\(m\)-approximation of
  \(\varphi\) (in \(T'\))} if \(\varphi'\) belongs to a theory \(T'\) s.t. \(T'\preceq T\), \(m\models \varphi'\) and \(\varphi'\Rightarrow\varphi\).
\end{definition}

An \(m\)-approximation has the following properties, which will be of use for us later on:\BCFV{TODO: add proofs in appendix?}
\begin{proposition}[\(m\)-approximation transitivity]\label{prop:m-approx-transitivity}
     If $\varphi''$ is an $m$-approximation of $\varphi'$ in $T''$ and $\varphi'$ is an $m$-approximation of $\varphi$ in $T'$, then $\varphi''$ is an $m$-approximation of $\varphi$ in $T''$.
\end{proposition}
\begin{proposition}[\(m\)-approximation conjunction closure]\label{prop:m-approx-conjunction}
    If $\varphi'_1$ is an $m$-approximation of $\varphi_1$ in $T'$ and $\varphi'_2$ is an $m$-approximation of $\varphi_2$ in $T'$, then $\varphi'_1\wedge\varphi'_2$ is an $m$-approximation of $\varphi_1\wedge\varphi_2$ in $T'$.\BCFV{not exactly closure, though}
\end{proposition}

The problem of \textit{model-guided approximation} (MGA) with respect to two theories $T,T'$ s.t. $T'\preceq T$ is now defined as: given a formula
$\varphi$ in $T$ and a model $m$ of $\varphi$, find an \(m\)-approximation of $\varphi$ in $T'$.  
In the sequel, we will often use the abbreviated phrase ``model approximation'' instead of the full ``model-guided approximation'', which is a mouthful.

%%% Local Variables:
%%% mode: latex
%%% coding: utf-8
%%% TeX-master: "main"
%%% TeX-PDF-mode: t
%%% TeX-engine: default
%%% End:
\section{Solving the MGA Problem}\label{sec:mbua-comp}
In this section, we focus on how to solve the model approximation problem with
respect to $T$ and $T'$.  For the remainder of this section\BCFV{of the paper?},
we fix $\varphi$ and $m$ to be the inputs of the problem ($\varphi$ is a formula
in $T$, $m$ is a model of $\varphi$).

A useful first step in solving this problem is via a reduction to the special
case where $\varphi$ is a product term, \ie a conjunction of literals.  We refer
to this case as the \textit{product model-guided approximation} (PMGA) problem.
Such a reduction is useful since it simplifies the problem without depending on
the particular $T$ and $T'$ in question.

In the literature, a product term $P$ s.t.\@ $P\Rightarrow\varphi$ is called an
\textit{implicant} of $\varphi$.  For our purposes, we add the notion of an
\textit{$m$-implicant} of $\varphi$, which is an implicant of $\varphi$ that is
satisfied by $m$.  Thus, by definition, an $m$-implicant of $\varphi$ is an
$m$-approximation of $\varphi$ in $T$\BCFV{I would like to remove the "in T"
  part here}, which is a product term.  Since $m$-approximation is transitive
(\cref{prop:m-approx-transitivity}), reducing the MGA problem of $\varphi$ and
$m$ to a PMGA problem can be done by simply replacing $\varphi$ with one of its
$m$-implicants.  We explain how to extract an $m$-implicant of $\varphi$ in~\cref{appendix:m-implicant}.

In the following, we focus on two instances of the PMGA problem and present
specialized algorithms for them.  \BC{removed a reference to appendix c - should we remove this appendix?}
% Our motivation for choosing these instances, as well as a discussion on possible other choices, are presented in~\cref{appendix:additional-theories}.
% The first subsection is focused on the instance where $T=\Tmia$ and $T'=\Tic$.
% In the second subsection, we define two theories, $\Tamia$ and $\Taic$, which
% extend $\Tmia$ and $\Tic$ with arrays and function calls, resp.  We then solve
% the instance where $T=\Tamia$ and $T'=\Taic$.  Note that $\Tic$ indeed
% restricts $\Tmia$ according to \cref{def:theory-restriction}.

\subsection{Approximating the Theory of Linear Integer Arithmetic with Non-Linear
  Multiplication Using the Theory of Intervals }\label{subsec:rules}

The first instance we present a solution for is computing the PMGA problem for
$T=\Tmia$ and $T'=\Tic$.  That is, given a product term $P$ in $\Tmia$ and a
model $m$ of $P$, our goal is to find a formula $\varphi'$ in $\Tic$
s.t. $\varphi'$ is an $m$-approximation of $P$.  To do that, for each literal
$l$ in $P$, we find an $m$-approximation of $l$, denoted $\xi'$ (note that
$\xi'$ is not necessarily a literal).  Then, we conjoin all $\xi'$s back
together to form the formula $\varphi'$.  From~\cref{prop:m-approx-conjunction}, $\varphi'$ that is created this
way is indeed an $m$-approximation of $P$.

In order to find an $m$-approximation for a $\Tmia$ literal $l$, we use a
rule-based approach.  \Cref{fig:integer-rules} shows inference rules to
transform ${\Tmia}$ literals into ${\Tic}$ formulas.  We assume, without loss of
generality, that literals have the form $t \leq c$ where $c$ is a constant and
$t$ has no constant terms (addends).  Integer inequalities can always be
normalized to this form.  In \cref{rule:int-add}, addition is modeled by
proportionally dividing the ``slack'' between the value of the constant and the value of the left-hand side in $m$ among the variables. For
example, given \(x_1+x_2+x_3\le 11\) and a model
\(m=\{x_1\mapsto 1, x_2\mapsto 2, x_3\mapsto 3\}\), we are distributing \(11-6\)
amongst three variables, and can approximate that \(x_1\le 1+2\),
\(x_2\le 2+2\), and \(x_3\le 3+1\).  Multiplication by a constant
(\cref{rule:int-mult-const}) is fairly simple, and is done by dividing the
right-hand~side of the constraint by the constant, only taking care to adjust
for negative constants correctly.  With multiplication of variables, we only use
the calculated sign of the result to provide a range in which the sign is
consistent with the model, shown in \cref{rule:int-mult-1,rule:int-mult-2}.  If
the product is positive, then factors are allowed to expand toward zero, and if
it is negative---away from zero.  For example, if we have a constraint
$x_1\cdot x_2 \leq -42$ and a model $m=\{x_1\mapsto 5,x_2\mapsto -9\}$, then
from \cref{rule:int-mult-2} we get $x_1 \geq 5$, $x_2 \leq -9$.

The rules as shown in \cref{fig:integer-rules} follow some conventions to aid
readability.\BC{comment from reviewer: make it clear that they should be read
  bottom-up} All formulas are assumed to be in canonical form: all the variables
are on the left-hand side with only a constant,~\(c\), on the right-hand side,
and inequalities are normalized into less-than forms, \ie
\(x\ge c\Rightarrow -x\le -c\). \(t_i\) denotes terms in the formula (not always
literals). \(\Eval{t_i}\) is the matching values per the model
\(m\). \(\Sign(x)\) gives the sign of \(x\), \ie
\(\frac{x}{|x|}\).

\begin{figure*}[t]
  \hspace{-1em}
  \vspace{1em}
  \begin{minipage}{0.5\textwidth}
  \begin{equation}
    \begin{adjustbox}{max width=\textwidth-3em}
    \begin{prooftree}
      \hypo{t_i\le \Eval{t_i} + \Portion\left(c-\sum_i \Eval{t_i},k,i\right)}
      \hypo{i=1\isep k}
      \infer2{t_1+\dots+t_k \le c}
    \end{prooftree} \label[rule]{rule:int-add}
      \end{adjustbox}
  \end{equation}
  \end{minipage}
  \begin{minipage}{0.5\textwidth}
  \begin{equation}
  \begin{adjustbox}{max width=\textwidth-3em}
    \begin{prooftree}
      \hypo{\prod_i \Eval{t_i} < 0}
      \hypo{\Sign\left(\Eval{t_i}\right)\cdot t_i\ge \left|\Eval{t_i}\right|  }
      \hypo{i=1\isep k}
      \infer3{t_1\cdots t_k \le c}
    \end{prooftree} \label[rule]{rule:int-mult-2}
    \end{adjustbox}
  \end{equation}
  \end{minipage}
  
  \vspace{-1em}\hspace{-1em}
  \begin{minipage}{0.4\textwidth}
  \begin{equation}
  \begin{adjustbox}{max width=\textwidth-3em}
    \begin{prooftree}
      \hypo{\Sign(a)\cdot t_1 \le \Sign\left(a\right)\cdot \left(c//a\right)}
      \infer1{a\cdot t_1 \le c}
    \end{prooftree} \label[rule]{rule:int-mult-const}
  \end{adjustbox}
  \end{equation}
  \end{minipage}
  \begin{minipage}{0.6\textwidth}
  \begin{equation}
  \begin{adjustbox}{max width=\textwidth-3em}
    \begin{prooftree}
      \hypo{\prod_i \Eval{t_i} \ge 0}
      \hypo{0\le\Sign\left(\Eval{t_i}\right)\cdot t_i\le\left|\Eval{t_i}\right|}
      \hypo{i=1\isep k}
      \infer3{t_1\cdots t_k \le c}
    \end{prooftree} \label[rule]{rule:int-mult-1}
  \end{adjustbox}
  \end{equation}
  \end{minipage}
  
  \vspace{0.5em}
  \scalebox{0.8}{Auxiliary notation:\qquad
  $
\begin{array}{r@{~}c@{~}l@{\quad}r@{~}c@{~}l}
  \Portion\left(n,k,i\right) & \triangleq &
  \left\lfloor\frac{n}{k}\right\rfloor +
  \begin{cases}
    1,& i\le n \mod k \\
    0,& \text{otherwise}
  \end{cases}
      %\\
      &
          x // y &\triangleq&
          \begin{cases}
            \left\lfloor \nicefrac{x}{y}\right\rfloor, & y> 0 \\
            \left\lceil \nicefrac{x}{y}\right\rceil, & y< 0
          \end{cases}
\end{array}
$}
  \caption{Rules for strengthening (transforming) integer terms to interval
    terms}
  \label{fig:integer-rules}
  \vspace{-1em}
\end{figure*}

\begin{example}
  As a demonstration of the application of these rules, recall
  \cref{fig:intro-example}.  In the addition node, we demonstrate the usage of
  \cref{rule:int-add}. The amount of ``slack'' to be shared between the child
  nodes is \(c-\sum_i \Eval{t_i}=7-12+10=5\), as noted in the value below the
  node. \(5\) cannot be evenly divided into two parts, therefore \(\Portion\)
  assigns \(3\) to the first child and \(2\) to the other.  This gives
  $x\leq 15$ and $-5y \leq -8$.  The multiplication node in this example
  represents multiplication by constant $-5$, therefore
  \cref{rule:int-mult-const} is applied. The constant is negative, so the
  operator is reversed, and since \(\left\lceil\nicefrac{8}{5}\right\rceil\) is
  \(2\), we obtain \(y\ge 2\).
\end{example}
\subsection{Adding Arrays and Uninterpreted Functions}\label{sec:arrays}
In this section, we solve the PMGA problem of $P$ and $m$ with respect to two integer theories with arrays and uninterpreted functions, $\Tamia$ and $\Taic$.
The theory of arrays, uninterpreted functions, and linear integer arithmetic with multiplication, $\Tamia$, is defined
via the grammar shown in \cref{fig:Tamia-grammar}, where $\IntConst$ is an integer constant, $v$ is an int variable symbol, $a$ is an array variable symbol, and $f_i$ are uninterpreted function symbols $\not\in\Sigma(\Tmia)$.
\footnote{The arity of functions is restricted to 1 for simplicity of presentation; 
an extension of our algorithm to functions with arbitrary arity is straightforward.
% Our implementation supports functions with any arity.}.
}

\begin{figure}[t]
    \begin{subfigure}[b]{0.5\textwidth}
    \centering
\begin{adjustbox}{max width=\textwidth}
$\displaystyle % Workaround for adjustbox, \[ \] fails
\begin{array}{r@{~}c@{~}l}
\varphi & ::= &  \lnot\varphi ~|~ \varphi\land\varphi ~|~ \varphi\lor\varphi ~|~ Atom \\
Atom & ::= & t < t ~|~ t \leq t ~|~ t > t ~|~ t \geq t ~|~  t = t ~|~ t \neq t  \\
     & ~|~ &   s = s ~|~ s \neq s \\
t & ::= & \IntConst ~|~ v ~|~ t + t ~|~ t - t ~|~ t\cdot t ~|~ f_i(t) ~|~ select(s,t) \\
s & ::= & a ~|~ store(s,t,t)
\end{array} %
$ %
\end{adjustbox}
    \caption{$\Tamia$}
    \label{fig:Tamia-grammar}
    \end{subfigure}
    \hfill
    \begin{subfigure}[b]{0.5\textwidth}
    \centering
\begin{adjustbox}{max width=\textwidth}
    $\displaystyle % same workaround as before
    \begin{array}{rcl}
\varphi & ::= & \varphi\land\varphi ~|~ Atom \\
Atom & ::= & \psi < \IntConst ~|~ \psi \leq \IntConst ~|~
  \psi > \IntConst ~|~ \psi \geq \IntConst \\
\psi & ::= & v ~|~ select(a,t) ~|~ f_i(t)\\
t & ::= & \IntConst ~|~ v ~|~ t + t ~|~ t - t ~|~ \IntConst\cdot t ~|~ f_i(t) ~|~ select(a,t) \\
s & ::= & a ~|~ store(s,t,t)\\
\end{array}
    $
\end{adjustbox}
    \caption{$\Taic$}
    \label{fig:Taic-grammar}
    \end{subfigure}
    \caption{Grammars for theories used}
    \label{fig:array-grammars}
\end{figure}

Integer variables are mapped to the $\IntDomain$ domain, array variables are mapped to the $\IntDomain\to\IntDomain$ domain (both the indices and the values are integers), and function symbols are also mapped to the $\IntDomain\to\IntDomain$ domain (both the argument and the return value are integers).
Except for the variable and function symbols, all symbols are interpreted using the standard interpretation.
Specifically, the interpretation of $\aselect(s,i)$ intuitively means querying the value of array $s$ in index $i$, and the interpretation of $\astore(s,i,e)$ intuitively means returning a fresh array which results from cloning $s$ and then storing the value of $e$ in index $i$.
We refer to terms of the form $f_i(\_)$, $\aselect(\_,\_)$, and $\astore(\_,\_,\_)$ as \textit{function application}, \textit{select}, and \textit{store terms}, resp.

To under-approximate formulas in $\Tamia$, we use the theory $\Taic$ as shown in \cref{fig:Taic-grammar}, which allows for intervals over integer variables, function call terms, and select terms where the first argument is an array variable.

\begin{comment} % ok to remove?
\color{gray}
no function calls, just arrays.

We refer to terms of the form $f_i(t)$ and $select(s,t)$ as \textit{complex}\BC{better name?} terms.
We denote the value of an integer variable $v$ in a $\Tmia$-structure $s$ by $\Eval{s}{v}$.

A model of the formula gives interpretation to arrays as functions from indices to values. E.g., $m[a]$ is a function $\Int\to\Int$ for an array $a$ of sort \texttt{(array int int)}.
Function symbols are also interpreted as functions, in the normal way.

Define syntatic replacement operator. Sqsubseteq relation.
\end{comment}

For now, let us ignore the presence of literals of the form $s=s$ or $s\neq s$ in $T$ (that is, assume the input formula contains no array comparison).
This is done for simplification of presentation; later we will show how to lift this restriction.
For similar reasons, since a function application term $f(i)$ is handled similarly to the select term $select(f,i)$, we assume, without loss of generality, that only the former\BC{should be the latter?} are present.
We consider the exact identification of such terms to be an implementation detail.\BC{What does this sentence mean?}

The derivation of an $m$-approximation for $P$ (a product term in $\Tamia$) in $\Taic$ now proceeds in multiple stages (explained below):
\begin{itemize*}
    \item Elimination of $\aselect$-$\astore$;
    \item Atomic grounding;
    \item Applying the PMGA procedure for $\Tmia$ and $\Tic$ (\cref{subsec:rules});
    \item Un-grounding the resulting formula.
\end{itemize*}

To ease the presentation of the rest of the section, we introduce a few notations.
Since several stages make use of syntactic replacement, we use $t\synin \varphi$ to denote that the term $t$ appears syntactically as a sub-term in the formula $\varphi$;
and $\sreplace{\varphi}{t}{t'}$ for the formula obtained by syntactically replacing every occurrence $t\synin\varphi$ with $t'$.

% The\BC{first attempt - explaining the procedure explicitly. I don't like it becuase it's an inlined algorithm. I thought it would be straight-forward to exlpain but there are too many details} first stage is a preprocessing step in which we use the model $m$ to remove all sub-terms of the form $\aselect(\astore(\_), \_)$ from $P$:
% For each literal $l\in P$ s.t. there exist $t\sqsubseteq l$ of the form $t=\aselect(\astore(s,t_i,t_e),t_j)$, we syntactically replace $t$ with $t_e$ in $l$ if $t_i=t_j$ is true in $m$; otherwise, we replace $t$ with $\aselect(x,t_j)$.
% In the former case, we also add the literal $t_i=t_j$ to $P$, and in the later case we add $t_i\neq t_j$. 
% The procedure is then recursively called for all new terms in $P$.
% When this stage is complete, all select terms $t\sqsubseteq P$ are of the form $\aselect(a,t)$, where $a$ is an array variable symbol.
% with conditional terms (\textit{ite}) such that the array arguments of select-terms are single array variable symbols.

The first stage\BC{if we want to save space we can treat this as a known procedure and also lose the syntactic replacement notation} is a preprocessing step in which we use the model $m$ to remove all $\aselect$-$\astore$ sub-terms (\ie terms of the form $\aselect(\astore(\_), \_)$) from $P$:
For every literal $l\in P$ s.t.\@ there exists $t=\aselect(\astore(s,t_i,t_e),t_j)\sqsubseteq l$, we remove $l$ from $P$, construct an equivalent formula $\varphi_l$ without $\aselect$-$\astore$ sub-terms (as shown below), find an $m$-implicant $P_l$ of $\varphi_l$ using the method described in~\cref{appendix:m-implicant}, and conjoin $P$ with $P_l$. The formula $\varphi_l$ is constructed recursively. 
Initially $\varphi_l$ is: 
$$\big( (t_i=t_j) \land \sreplace{l}{t}{t_e} \big) \lor
\big( (t_i\neq t_j) \land \sreplace{l}{t}{\aselect(s,t_j)} \big)$$
Then, the term $\aselect(s,t_j)$ is recursively replaced likewise, as long as $s$ is a $\astore$. The second stage is the grounding procedure, whose purpose is to connect the semantics of $\Tamia$ formulas with that of $\Tmia$, with the aim of reducing the problem.

%%%%%%%%

%Projection applied to formulas
%MIA-projection of a formula $\varphi$

We define an injective \emph{grounding function} $t \mapsto v_t$ over terms of $\Tamia$ that assigns to any term $t$ of sort $S$ a unique variable symbol $v_t$ of sort $S$.
Given a formula $\varphi\in\Tamia$, we define its \emph{atomic grounding} $\agphi$
that is obtained from $\varphi$ by (syntactically) replacing 
every select-term $t\sqsubseteq\varphi$ with $v_t$.

% Projection applied to structures
Similarly, we define \emph{atomic grounding on structures}, for structures of $\Tamia$.
A structure $\agm$ is obtained from a $\Tamia$-structure $m$ by elision of all array interpretations and introduction of appropriate interpretations $\agm[v_t] = m[t]$.
Clearly, doing that for every possible $t$ will lead to infinite structures;
therefore, atomic grounding of structures is always done with respect to some set $S$ of terms, such that $v_t$ is assigned only for $t\in S$. We denote this by $m\overset{S}{\mapsto}\agm$.
In particular, having received an input formula $\varphi$, we fix $S$ to be the set of all select-terms occurring in $\varphi$.

It is worth noting that the mapping $\varphi\mapsto\agphi$ is readily invertible,
while $m\mapsto\agm$---not necessarily.
For example, the model $\agm^{\star} = \langle i\mapsto 0, j\mapsto 0, v_{\aselect(a,i)}\mapsto 1, v_{\aselect(a,j)}\mapsto 2\rangle$
has no model $m$ such that $\agm=\agm^{\star}$;
because any $\agm$ that maps $i$ and $j$ to the same value must also map $v_{\aselect(a,i)}$ and $v_{\aselect(a,j)}$ to the same value.

The process of generating an $m$-approximation of $\varphi \in \Tamia$ is as follows:
first, construct $\agphi$ and $\agm$.
Subject to the language restrictions above, $\agphi\in\Tmia$.
Then, generate an $\agm$-approximation of $\agphi$ as a new formula $\agphi'\in\Tic$.
Finally, construct the corresponding $\varphi'\in\Taic$ of which $\agphi'$ is an atomic grounding.
We will now show that $\varphi'$ is indeed an $m$-approximation of the original $\varphi$.

%Lemma that connects $\tilde{m}$ with $\tilde{phi}$.
\begin{lemma}[grounding fidelity]\label{lemma:grounding_fidelity}
Given $\varphi,m$ over a theory with arrays, $m\overset{S}{\mapsto}\agm$, for some $S$ that contains (at least) all select-terms occurring in $\varphi$. Then
$m\models\varphi \iff \agm\models\agphi$.
\end{lemma}

%A proof of this lemma can be found in~\cref{sec:proofs}
%Commutativity diagram with two directions.
%(+ Theorem)
\begin{proof}
In~\cref{appendix:grounding-proof}.
\end{proof}

Based\BC{the entire proof can move to the appendix, if necessary} on the lemma and the correctness of the $\Tmia$ model-based approximation procedure of \cref{subsec:rules}, we can now establish a correct approximation of $\Tamia$, based on the stages above.
From the lemma, $\agm\models\agphi$, and since
$\agphi'$ is an $\agm$-approximation of $\agphi$, it follows that $\agm\models\agphi'$.
Recall that the variables of $\agphi'$ are (a subset of) the variables of $\agphi$ and as a consequence, the select-terms occurring in $\varphi'$ also occur in $\varphi$, hence are contained in $S$.
We can therefore apply the lemma in the opposite direction, and obtain $m\models\varphi'$.

The second property that $\varphi'$ must satisfy is that $\varphi'\Rightarrow\varphi$.
The proof is very similar:
let $m'\models\varphi'$ be some model of it,
and construct $\agm'$.
From the lemma, $\agm'\models\agphi'$.
Again, $\agphi'$ under-approximates $\agphi$, hence $\agm'\models\agphi$;
thus from the lemma, $m'\models\varphi$.

The astute reader may observe that the procedure above, while fulfilling the requirements for being a model-based approximation in $\Taic$,
is somewhat unsatisfying because $\Taic$ formulas are not as easy to sample as $\Tic$.
Sampling the individual array elements independently from their corresponding intervals may lead to clashes; \eg if $i$, $j$, $\aselect(a,i)$, and $\aselect(a,j)$ are all constrained to some interval $[c,d]$,
then an assignment of $i = j$ and $\aselect(a,i)\neq \aselect(a,j)$ is inconsistent and does not yield a valid structure.
One possible course of action is to detect clashes as soon as they occur and restart the sampling procedure.
But this may lead to a large number of false flags and wasted computations, which was the whole point of a sampling algorithm to avoid in the first place.
A better alternative would be to adjust the construction of $\varphi\in\Taic$ in a way that avoids these clashes.

To do that, we first create two sets of literals, $L_{=}$ and $L_{\neq}$, which constrain the indices to a fixed aliasing configuration, namely the one exhibited by $m$.
\[
\fontsize{7pt}{7pt}\selectfont
\renewcommand\arraystretch{1.5}
\begin{array}{l@{}l}
L_{\neq}=\{& t_1\neq t_2 ~|~ \exists a.\, \aselect(a,t_1),\aselect(a,t_2)\sqsubseteq P ~\wedge~ m[t_1]\neq m[t_2]\}
\\
L_{=}=\{& t_1=t_2,\aselect(a,t_1)=\aselect(a,t_2) ~|~ \\
& \exists a.\aselect(a,t_1),\aselect(a,t_2)\sqsubseteq P ~\wedge~ m[t_1]=m[t_2]\}
\end{array}
\]

Adding these literals to $P$ will make sure that the same array accesses that are aliased in $m$ are aliased in any model of $\varphi'$, and no others.

Finally, we handle array equality $a_1 = a_2$ (and its negation) via a similar preprocessing. Since the details are somewhat technical, they are deferred to  \cref{appendix:array-eq}.

\section{Sampling Using Model-Guided Approximation}\label{sec:sampling}

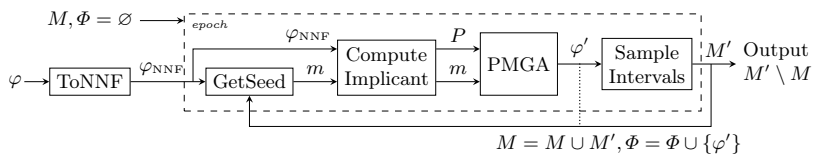
\begin{figure}[t]
\centering
 \resizebox{0.9\textwidth}{!}{
 \begin{tikzpicture}[>=stealth,
    block/.style={rectangle,draw,align=center},
    pt/.style={coordinate}
  ]
  \node(start)[inner sep=1pt] {$\varphi$};
  \node(nnf)[right=4mm of start,block] {ToNNF};
  \node(start-m)[above=5mm of nnf] {$M,\Phi=\varnothing$};
  \node(epoch-start)[right=10mm of nnf,pt] {};
  \node(seed)[right=12mm of nnf,block] {GetSeed};
  \node(implic)[right=7mm of seed.south east,
                anchor=south west,block]
               {Compute\\Implicant};
  \node(pgma)[right=7mm of implic.south east,
              anchor=south west,block,
              minimum height=10mm]
             {PMGA};
  \node(blk)[right=3.5mm of pgma,pt] {};
  \node(milk)[right=7mm of pgma,block] {Sample\\Intervals};
  \node(end-2)[right=3mm of milk,pt] {};
  \node(out)[right=4mm of end-2,align=left] { Output\\$M'\setminus M$ };
  
  \draw[->] (start) -- (nnf);
  \draw[->] (nnf) -- 
      node[above]{$\varphi_{\textrm{\tiny NNF}}$}
      (epoch-start) -- (seed);
  \draw[->] (seed) -- node[above]{$m$} 
      (seed -| implic.west);
  \draw[->] (implic.east |- seed) -- node[above]{$m$}
            (pgma.west |- seed);
  \draw[->] (pgma) -- node[above]{$\varphi'$} (milk);
  \draw[->] (milk) -- node[above,pos=0.6]{\small$M'$} (out);
  
  \node(d)[pt,above=3mm of implic.west]{};
  \draw[->] (epoch-start) |- 
      node(lphi)[above,pos=0.9]
      {$\varphi_{\textrm{\tiny NNF}}$} (d);
  
  \node(epoch) 
       [fit=(epoch-start)(lphi.12)(seed)(implic)(pgma)(milk),
        draw,dashed,
        inner xsep=4pt,inner ysep=7pt] {}; 

  \node[at=(epoch.north west), anchor=north west] {\tiny \textit{epoch}};

  \draw[->] (start-m) -- (start-m -| epoch.west);

  \draw[->] (implic.east |- d) --
             node[above]{\small $P$}
            (pgma.west |- d);

  % back-edge with dotted edge from phi'
  \draw[line cap=round, line width=0.65pt,
    dash pattern=on 0 off 1.5pt] (blk) -- +(0, -10mm);
  \draw[->] (end-2) -- +(0,-10mm) -|
    node[below,pos=0.1]{
      $M = M\cup M',
      \Phi = \Phi\cup\{\varphi'\}$}
  (seed);
  %\draw[->] (end-2) -- +(0,-10mm) -| 
  %   node[below,pos=0.25]{$M = M\cup M'$} (epoch.-12);
\end{tikzpicture}%
 }%
 \caption{Workflow of the \(\texttt{MeGASample}\) sampling procedure }
 \label{fig:workflow}
\end{figure}

In this section, we present an algorithm, \(\texttt{MeGASample}\), for sampling $\Tmia$ and $\Tamia$ formulas, using model-guided approximation to the theories $\Tic$ and $\Taic$, respectively.
An outline of the algorithm is presented in \cref{fig:workflow}.
Similar to~\cite{Dutra:2018}, the sampling process is \textit{epoch-based}:
it obtains an initial model with the help of a solver, and then utilizes it to create a set of distinct models in an efficient manner.
This process is repeated iteratively, where each such iteration is called \textit{an epoch}.
The initial model of each epoch is called the \textit{seed} of the epoch.

As a preliminary step, the input formula $\varphi$ is transformed into negation~normal~form (NNF).
% , since this is a prerequisite to \texttt{ComputeImplicant} in subsequent iterations.
% Conversion to NNF is done in the standard way, by propagating negation inwards using De-Morgan's laws, without increasing the number of literals in the formula~\cite{Book2001:Degtyarev}.
Then, the epoch loop begins, collecting models in the set $M$ (initialized to $\varnothing$).
First, the seed of the epoch, $m$, is obtained using a \texttt{GetSeed} procedure.
% , which, following~\cite{Dutra:2018}, picks a random assignment and chooses the model of $\varphi$ that is closest to it using a MAX-SMT query.
%\SI{this note now appears, in different words, below}
%Note that, in addition to $\varphi_{\mathrm{NNF}}$, this procedure also gets the underapproximated formula $\varphi'$ of the previous epoch, which is sometimes used to guide the choice of the seed.
% Details of this procedure are provided below.
Then, model-guided approximation takes place.
As discussed in~\cref{sec:mbua-comp}, the first step is to find a product term $P$ that is an $m$-implicant of $\varphi_{\mathrm{NNF}}$ (and $\varphi$) in \texttt{ComputeImplicant}.
% , in order to reduce the MGA problem to a PMGA problem.\BC{make sure these initials are consistent in all places} 
Then, the procedure \(\texttt{PMGA}\) that is appropriate to the current theory is used to obtain $\varphi'$, an $m$-approximation of $P$ (and $\varphi$).
Finally, in \texttt{Sample intervals}, the interval formula $\varphi'$ (which is either in $\Tic$ or $\Taic$) is repeatedly sampled, and samples are saved in $M'$.
% by drawing from the interval domains specified therein, and a set $M'$ of models of it is returned.
Note that, the \(m\)-approximation property guarantees that every model of \(\varphi'\) is also a model of \(\varphi\).
Therefore, all models in $M'$ are output if they were not seen before (\ie not in $M$), and the set $M$ is updated accordingly.
In addition, the \(m\)-approximations $\varphi'$ are accumulated for future use by \texttt{GetSeed}.

We suggest two variants for the \texttt{GetSeed} procedure.
The first, following~\cite{Dutra:2018}, picks a random assignment and chooses the model of $\varphi$ that is closest to it using a MAX-SMT query.
The second, avoids a costly MAX-SMT query and uses SMT instead, but only after discrading all models of $\varphi'$ by adding $\neg\varphi'$ to the formula.
Unlike the random-based variant, constraints are accumulated between different calls to the solver, to enable blocking based on the entire history of epochs. 
Both variants aim at increasing the diversity of the seeds and covering new areas of the solution space.

With regard to \texttt{Sample Intervals}, for $T'=\Tic$ this procedure can be realized by repeatedly drawing a value at random from within the lower and higher bounds of each variable. 
For $T'=\Taic$, things get a little more complicated since we need to also sample array elements constrained by an interval.
As already mentioned in~\cref{sec:arrays}, clashes may occur between terms that are not syntactically identical but refer to the same array element (are aliased).
% We could use a generate-and-test approach on top of the $\Tic$ sampling procedure (reference implementation is shown in~\cref{alg:sampleaic}).
However, we have already shown that with the addition of equality and inequality constraints we can eliminate aliasing a-priori. 
We therefore use the $\Tic$ sampling procedure also for $\Taic$. 

The suggested implementations for \texttt{GetSeed} and \texttt{Sample Intervals} are suitable for the general case, where we have no information as to the form of the solution space nor the goal of sampling.
However, we believe that for a particular application such additional insight will be available and can be incorporated into these procedures in order to guide the search towards the more useful solutions.
We leave this to future work. \BC{This paragraph is a draft. we can remove it to go back to the page limit}

\begin{comment}
\paragraph{\MeGASampler}
Next, we present a sampler, called \MeGASampler, which instantiates the \(\texttt{MeGASample}^T_{T'}\) framework for the two pairs of theories $T=\Tmia,T'=\Tic$ and $T=\Tamia,T'=\Taic$ using the PMGA algorithms for these theories (\cref{subsec:rules} and \cref{sec:arrays}).

In addition, the \texttt{Sample Intervals} 
uses some local heuristics to decide how much to invest in the sampling of intervals based on their size.
% It does that by sampling the intervals in rounds, where in each round the number of samples is limited in proportion to the interval's size, and also by stopping early when the rate of unique new samples generated falls under a predefined limit.
% To do that, it performs \(n\) sampling rounds in which \(m\) randomly-drawn samples are produced using \(\texttt{Sample}_{T'}\). 
% After finishing each round, if the rate of unique new samples generated falls under a
% predefined limit, the sampling of this formula is stopped. 
% Additionally, one can heuristically adjust \(m\) and \(n\) based on properties of the formula

Further technical details of \MeGASampler are available in~\cref{appendix:technical-details}.
\end{comment}

\section{Evaluation}\label{sec:eval}
We have developed \MeGASampler as an open-source sampling tool based on the Z3
solver~\cite{DeMoura:2008}. In this section, we empirically evaluate
\MeGASampler with either blocking or random initial assignments, and compare it
against a variant of state-of-the-art sampler \SMTSampler~\cite{Dutra:2018}
which was ported from bit-vector logic to integer logic. The latter was
constructed by modifying the original implementation as follows. \SMTSampler is
built upon three operations: transforming a seed into a set of satisfying
conditions, finding similar solutions by negating one of these conditions, and
combining the resulting solutions to generate solution candidates. Porting to a
different logic requires translating these operations to fit that logic. For
bit-vectors, \SMTSampler used conditions of the form ``bit \(i\) in vector \(v\)
is set'', and combined solutions by applying a bitwise~or operation to mutated
values. For the integer domain, we treated the entire assignment as a condition,
\ie ``integer \(x\) is equal to value \(n\)'', and used addition as the integer
analogue for combining mutations.

We performed tests on benchmarks from SMT-LIB~\cite{Barrett:2021}, using the
problems in logics QF\_LIA, QF\_NIA\footnote{The ``interesting'' operation in
  $\Tnia$ is multiplication; most benchmarks in this directory are actually in
  $\Tmia$, which is supported by \MEGASampler.}, and QF\_ALIA.  Some of the
benchmarks represent real-world problems, including formulas used for software
verification, \eg from the \textsc{AProVE}~\cite{Giesl:2004} and
\textsc{VeryMax}~\cite{Borralleras:2017} termination analysis tools. Other
benchmarks are synthetic, designed to stress SMT solvers.

Benchmarks deemed inappropriate for the evaluation were discarded, including
those marked as unsatisfiable or unknown and benchmarks for which at least 100
samples were not gathered by any tested technique. As we are not evaluating SMT
solvers, benchmarks for which finding a single solution took over one minute
were also discarded. After applying these criteria, we followed the methodology
of \cite{Dutra:2018} and randomly selected 15 benchmark files as a
representative sample of each directory, %with a sufficient number of files,
in order to keep the experiments tractable. We consider this reasonable as
benchmarks from the same directory tend to be similar in nature. In total, our
evaluation set consisted of 28 benchmark directories (420 files), 4 in QF\_NIA,
1 in QF\_ALIA, and the rest in QF\_LIA.

% Metric
As noted in previous work~\cite{Dutra:2018}, the number of unique solutions
generated is an incomplete metric, which may not represent the samples' coverage
of the solution space.
% However, coming-up with an appropriate metric for the evaluation of a
% general-purpose SMT sampler is very challenging.  First, because it is not
% always clear how to define and measure coverage for a theory $T$, especially
% over an infinite domain.  Second, because coverage may not always be the
% appropriate metric, as different applications require different things from
% the solution space.
The authors proposed an alternative metric, which measures coverage statistics
of the internal nodes of the abstract syntax tree~(AST) of the (bit-vector)
formula: each node of sort Bool represents 1 bit, and each node of sort
bit-vector represents 64 bits. Each such bit is considered covered if it has
received both 1 and 0 among the set of samples. The coverage metric is then the
ratio of covered bits to total bits.
% hypothesizing that the inner nodes of the formula abstract syntax tree~(AST)
% offer higher-level information than just the leaf variables nodes.
The intuition being that if one were to synthesize a circuit that takes as
inputs assignments to the variables of the formula and produces a Boolean output
of \textit{True} or \textit{False} indicating whether the formula is satisfied,
then the coverage metric would be equivalent to the coverage of internal wires
in this circuit, when exercised by the generated solutions. Therefore, this
metric serves as a good proxy to coverage for formulas that encode hardware
systems, \eg in CRV~\cite{CRV}, and a good proxy to path coverage for formulas
that encode software systems, \eg in bounded model checking~\cite{cbmc}.

We have ported this metric from bit-vector logic to integer logic by considering
only the lower 64 bits of each arbitrary-size integer. Note that the maximum
score using this metric is not necessarily 100\%, because, naturally, some bits
are restricted by the input formula and can only take one of the values.
Therefore, we have normalized it and calculated it instead as the ratio of
covered bits to total bits covered by at least one of the methods in the
evaluation.

% Setting
All experiments were run on a 64-core (128-thread) AMD~EPYC~7742 based server
with 512GiB of memory.  Each execution utilized a single core, until either a
time limit of 15 minutes was reached, or 25M unique samples were generated.
% Additionally, each epoch was limited to 10 minutes, to prevent the samplers
% from getting stuck on a single seed.  Each experiment was repeated using
% different parameter values for the \texttt{Exploit$\varphi$'} procedure, which
% controls the sampling of intervals (\cref{appendix:technical-details}); we
% attempted the fixed-samples variant, with different values for the samples
% limit $x$, and the relative-samples variant, with different values for the
% coefficient $c$ and the rate $r$.\BC{I don't really know what I'm saying here
% :) revisit appendix- then this.}

\begin{table*}[t]
    \centering
    \newcommand\bx[1]{\textbf{#1}}
    \begin{adjustbox}{width=\textwidth,center=\textwidth}
    %\begin{threeparttable}
    \begin{tabular}{l | c c | r r r | r r r| r | r r r}
        \toprule
        &  & & \multicolumn{3}{c|}{Coverage (\%)} & \multicolumn{3}{c|}{Epochs} & \multicolumn{1}{c}{\#SMT} & \multicolumn{3}{|c}{Unique Solutions} \\
        Benchmarks & vars & depth & \MEGA & \MEGAb & \SMTint & \MEGA & \MEGAb & \SMTint & \SMTint & \MEGA & \MEGAb & \SMTint \\
        \midrule

        % benchmark a_ints b_depth c_coverage_MeGA c_coverage_MeGAb c_coverage_SMT d_epochs_MeGA d_epochs_MeGAb d_epochs_SMT e_smtcalls_SMT f_solutions_MeGA f_solutions_MeGAb f_solutions_SMT
QF\_ALIA/qlock2                                    & {     484} & {     233} & \textbf {   87.90} & {   20.40} & {   26.25} & \textbf {       3} & {       2} & {       2} & {      23} & {   12332} & \textbf {   14877} & {      20} \\
QF\_LIA/CAV2009-slacked\tnote{1}\;\,               & {      55} & {       5} & \textbf {   98.36} & {   66.11} & {   89.87} & {     143} & \textbf {     646} & {     247} & {    6746} & { 1647400} & \textbf { 1915995} & {  633047} \\
QF\_LIA/CAV2009\tnote{3}\;\,                       & {      26} & {       5} & {   67.38} & \textbf {   95.08} & {   79.98} & {     184} & {     996} & \textbf {    1585} & {   20793} & { 2201716} & \textbf { 3788132} & { 1075097} \\
QF\_LIA/bofill-sched-random\tnote{4}\;\,           & {     780} & {       6} & \textbf {   99.84} & {   83.95} & {   76.00} & {     628} & \textbf {     927} & {       4} & {     196} & {     627} & \textbf {     927} & {     112} \\
QF\_LIA/bofill-sched-real\tnote{5}\;\,             & {     576} & {       6} & \textbf {   99.94} & {   91.95} & {   81.74} & {     689} & \textbf {    1159} & {       3} & {     580} & {     688} & \textbf {    1158} & {     249} \\
QF\_LIA/convert                                    & {     768} & {    1339} & {   30.61} & {   27.71} & \textbf {   80.24} & \textbf {       7} & {       6} & {       2} & {      65} & \textbf {   90635} & {   76671} & {      53} \\
QF\_LIA/dillig                                     & {      31} & {      25} & {   42.20} & \textbf {   97.70} & {   53.57} & {     133} & \textbf {    1164} & {     769} & {   10551} & { 1729099} & \textbf { 3983851} & {  455503} \\
QF\_LIA/pb2010                                     & {    5842} & {       5} & \textbf {   75.64} & {   71.85} & {   51.37} & {   30099} & {     107} & \textbf {   42907} & {  128875} & {      53} & \textbf {     214} & {      27} \\
QF\_LIA/prime-cone                                 & {      10} & {       5} & {   96.89} & {   63.86} & \textbf {   97.00} & {    2705} & {    9313} & \textbf {   39360} & {  155484} & \textbf {13658385} & { 3906854} & {  613784} \\
QF\_LIA/slacks                                     & {      61} & {       5} & \textbf {   99.11} & {   69.50} & {   85.75} & {     135} & {     205} & \textbf {     383} & {   11005} & { 1923319} & \textbf { 2245506} & {  335504} \\
QF\_NIA/20170427-VeryMax                           & {     181} & {      13} & \textbf {   86.09} & {   54.47} & {   53.99} & {      13} & \textbf {     506} & {       2} & {      30} & {   81822} & \textbf { 1563188} & {   18698} \\
QF\_NIA/AProVE                                     & {      40} & {       6} & \textbf {   97.87} & {   71.57} & {   77.17} & {     714} & {    6016} & \textbf {   10053} & {   73190} & { 6478901} & \textbf { 6982253} & { 1394814} \\
QF\_NIA/leipzig                                    & {      92} & {       2} & {   59.55} & \textbf {   74.50} & {   54.22} & {      20} & \textbf {    1305} & {       3} & {      64} & {   96843} & \textbf { 1189523} & {    1207} \\
% coverage_MeGA=0.8149560974400868 solutions_MeGA=1884333.1151079137 coverage_MeGAb=0.7384711871197046 solutions_MeGAb=2272399.273381295 coverage_SMT=0.7706412352397715 solutions_SMT=559008.8393285371 coverage_megasampler=0.9421652649686911 solutions_megasampler=2078366.1942446043

        \bottomrule
    \end{tabular}
    %\end{threeparttable}
    \end{adjustbox}
    %\begin{tablenotes}
    %    \item[1] Short for \emph{CAV\_2009\_benchmarks/smt}
    %    \item[2] Short for \emph{20180326-Bromberger/more\_slacked/CAV\_2009\_benchmarks/smt}
    %    \item[3] Short for \emph{20180326-Bromberger/unbd-sage}
    %    \item[4] Short for \emph{bofill-scheduling/SMT\_random\_LIA}
    %    \item[5] Short for \emph{bofill-scheduling/SMT\_read\_LIA}
    %\end{tablenotes}
    \vspace{2.5pt}
    \caption{Results (averaged) over the benchmarks}
    \vspace{-10pt}
    \label{fig:mega-table}
\end{table*}

\subsection{Results}
\Cref{fig:mega-table} shows the results of the executions, broken down by origin
directory. For each directory of benchmarks, the columns, averaged over the benchmarks in that directory, are as follows: Number of variables, depth of the formula AST, the
computed coverage metric, number of epochs, number of
calls to the solver (only shown for \SMTint, for \MEGA and \MEGAb it is exactly
the number of epochs), and number of unique solutions. Columns titled
\MEGA\BCFV{Introduce these notations before using them on the previous sentence} report measurements for the random-based \MEGASampler, \MEGAb  the
blocking \MEGASampler, and \SMTint is \SMTSampler retrofitted for the integer
domain. Bold indicates the highest value in each rubric. Results are aggregated
per top-level directory; for a more detailed breakdown, see
\cref{appendix:results-full}.

\MeGASampler consistently produces more samples than \SMTint. Moreover, the
blocking technique is shown to be overall effective for increasing the number of
unique solutions.  Blocking does incur some overhead, which is why in some cases the
non-blocking variant was able to produce more samples within the time
frame. \MeGASampler also improves the coverage when considering the averages per
collection of 15 benchmarks. Here, blocking produces lower coverage even when it
produces more samples, because, while producing intervals that are
disjoint in successive epochs, it utilizes less randomization than the MAX-SMT-based
version.

The benchmarks in the \texttt{qlock2} and
\texttt{convert} categories were difficult to solve (possibly due to the large
AST depth), reflected in the low number of epochs for all methods shown.
Remarkably, \MEGA manages to extract tens of thousands of samples from a
single-digit number of seeds for these benchmarks, compared to tens of \SMTint.
Surprisingly, for \texttt{convert} it is \SMTint that manages to achieve the
greatest coverage, despite a low number of unique solutions, suggesting
that it moved further away from the seed using mutations.

For the two \texttt{bofill} categories, we see that for both \MEGA variants the number of samples is about
the number of epochs, meaning that MGA failed to generalize the seeds
obtained from the solver. In \SMTint the number of epochs is very small and the
number of solutions obtained is lower than the number of calls made to the
solver, \ie its seed expansion attempts were mostly
unsuccessful. One takeaway is that there are examples where generalization does
not work well but solving the formula isn't difficult, so maximizing the number of epochs is a good strategy. There were several categories where the number of epochs was significantly
larger in \MEGAb compared to the other two; we ascribe this result to the cost of
MAX-SMT calls, which aren't utilized by \MEGAb.

\begin{figure*}[t]
%\centering
\ifwithplots
  \resizebox{\textwidth}{!}{
    \begin{tikzpicture}[
  chart title/.style={font=\relsize{1.5}, inner sep=6pt}
  ]
  \begin{axis}[
        name=mega,
        ybar=0pt,
        bar width=0.2pt,
        enlarge x limits={abs = 10},
        xtick distance=100,
        ymode=log,
        ytick distance=10^1,
        ymajorgrids=true,
        xlabel={Benchmarks},
        ylabel={\#\,samples ratio},
        width=10cm,
        height=6cm]
        
    \addplot[draw=none, fill=blue]
      table[x expr=\coordindex,y=MeGA] {data/cactus-plot-num-samples/sorted-by-MeGA.dat.txt};
    \addplot[draw=none, fill=red!80!gray]
      table[x expr=\coordindex,y=MeGAb] {data/cactus-plot-num-samples/sorted-by-MeGAb.dat.txt};

  \end{axis}

  \node[chart title, at=(mega.north), anchor=north] 
    {{\color{blue}$\bullet$\,}{\MEGA}$/$\SMTint\hspace{.5em}%
     {\color{red}$\bullet$\,}{\MEGAb}$/$\SMTint\hspace{1em}~};

  \begin{axis}[
        name=megab,
        at=(mega.south east), anchor=south west,
        xshift=1cm,
        ybar,
        bar width=0.2pt,
        enlarge x limits={abs = 10},
        xtick distance=100,
        ymode=log,
        ytick distance=10^1,
        ymajorgrids=true,
        xlabel={Benchmarks},
        width=10cm,
        height=6cm]
        
    \addplot[draw=none, fill=purple]
      table[x expr=\coordindex,y=max] {data/cactus-plot-num-samples/sorted-by-max.dat.txt};

  \end{axis}

  \node[chart title, at=(megab.north), anchor=north] {max\{\MEGA,{\MEGAb}\}/\SMTint};

  \begin{axis}[
        name=mega,
        at=(megab.south east), anchor=south west,
        xshift=1.8cm,
        ybar=0.05pt,
        bar width=0.2pt,
        enlarge x limits={abs = 10},
        xtick distance=100,
        %ymode=log,
        %ytick distance=10^1,
        ymajorgrids=true,
        yticklabel={\pgfmathparse{\tick*100}\pgfmathprintnumber{\pgfmathresult}\%},
        xlabel={Benchmarks},
        ylabel={Coverage ratio},
        width=10cm,
        height=6cm]
        
    \addplot[draw=none, fill=blue]
      table[x expr=\coordindex,y=MeGA] {data/cactus-plot-coverage/sorted-by-MeGA.dat.txt};
    \addplot[draw=none, fill=red!80!gray]
      table[x expr=\coordindex,y=MeGAb] {data/cactus-plot-coverage/sorted-by-MeGAb.dat.txt};

  \end{axis}
  
  \node[chart title, at=(mega.north), anchor=north] 
    {{\color{blue}$\bullet$\,}{\MEGA}$-$\SMTint\hspace{.7em}%
     {\color{red}$\bullet$\,}{\MEGAb}$-$\SMTint};

  \begin{axis}[
        name=max,
        at=(mega.south east), anchor=south west,
        xshift=1cm,
        ybar,
        bar width=0.2pt,
        enlarge x limits={abs = 10},
        xtick distance=100,
        %ymode=log,
        %ytick distance=10^1,
        ymajorgrids=true,
        yticklabel={\pgfmathparse{\tick*100}\pgfmathprintnumber{\pgfmathresult}\%},
        xlabel={Benchmarks},
        width=10cm,
        height=6cm]
        
    \addplot[draw=none, fill=purple]
      table[x expr=\coordindex,y=max] {data/cactus-plot-coverage/sorted-by-max.dat.txt};

  \end{axis}

  \node[chart title, at=(max.north), anchor=north] {max\{\MEGA,{\MEGAb}\}$-$\SMTint};

\end{tikzpicture}
  }
\else
  \begin{tabular}{l}
  \includegraphics[width=4cm]{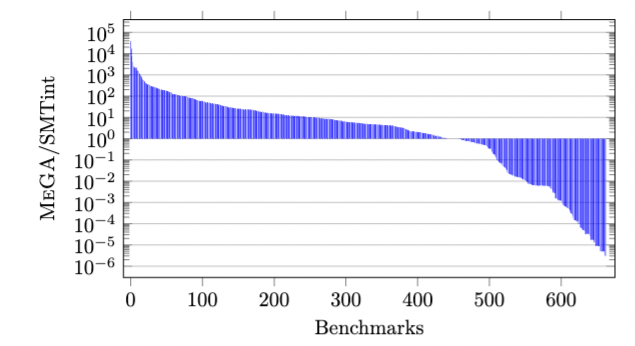}
  \includegraphics[width=4cm]{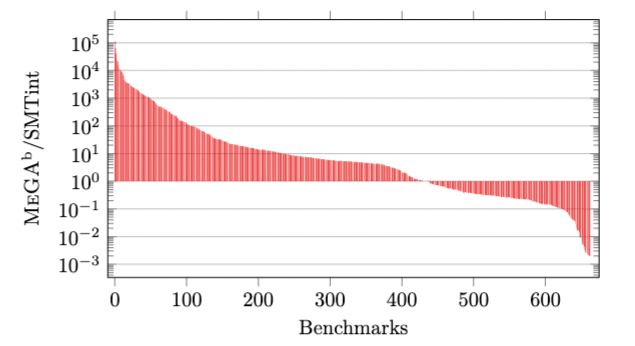}
  \includegraphics[width=4cm]{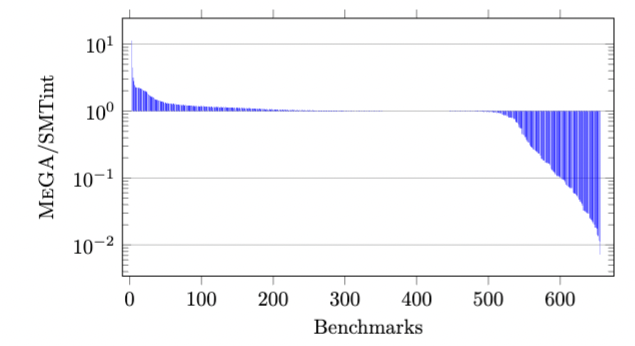}
  \includegraphics[width=4cm]{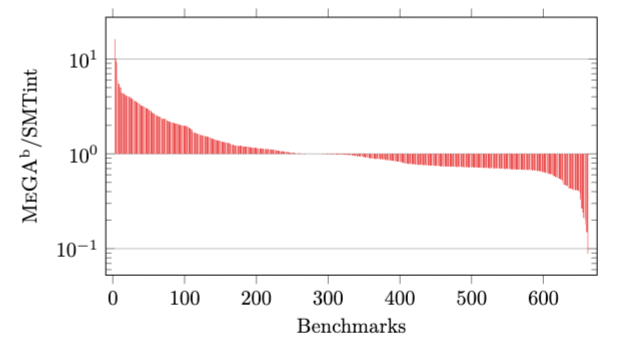}
  \\
(thumbnails used due to \texttt{\textbackslash withplotsfalse})
  \end{tabular}
\fi
  \caption{A comparison of the number of samples and coverage generated by \MeGASampler, relative to those from \SMTSampler.
  The plots show ratios per single benchmark, in ascending order,
  comparing each of two variants of \MeGASampler to \SMTSampler.
  }
  \label{fig:eval-plot-num-samples}
\end{figure*}
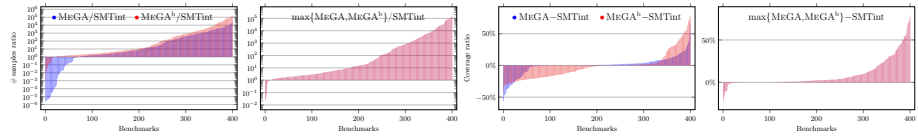

For a deeper look into the performance of the method, we perform a case-by-case
comparison between both variants of \MeGASampler and \SMTSampler.  We compute
the ratio between the number of samples generated by \MEGA or \MEGAb on each benchmark to \SMTint; then plot the results in descending
order on a logarithmic scale, on the same axis, as shown in
\cref{fig:eval-plot-num-samples}. We show the same for a ``virtual best'' where
the maximal value is compared with the baseline. The same process is repeated
for the coverage metric that was described earlier. Since coverage is a percentage metric, we plot the
difference on a linear scale.

In number of samples, the result of \MEGAb are similar to those
of the virtual best and support the previous conclusion that blocking is good
for increasing the sample volume. We also see that for coverage, the race is
largely inconclusive, but that the virtual best still offers significant
improvement over the baseline.

\begin{paragraph}{Limitations of the evaluation.}
  There are some notable limitations to the presented methodology. \SMTSampler
  was not designed to work on integers, and the modification may not be suitable
  or optimal. There are several parameters controlling sampling behaviors, \eg mutation depth in \SMTSampler or maximum number of
  samples per epoch in \MeGASampler. % which could impact the results.
  Different selections for these parameters, as well as other experimental
  parameters such as time limit and inputs, may impact the results.

  There may be other coverage
  metrics for estimating quality of samples, or other metrics that better
  characterize good sampling for particular uses.  However, we do not believe that there
  is any single metric that would be good for all applications. As future work,
  we plan to design an MGA-based goal-aware sampling algorithm that guides the
  exploration towards better solutions for a specific purpose.
\end{paragraph}
\section{Related Work}\label{sec:related}
There is a rich line of work on Markov-Chain Monte-Carlo (MCMC) sampling techniques in the statistics and operations research community~\cite{Glynn_Iglehart_1989,Hastings_1970,Liu_1996,Ozols_Roetteler_Roland_2013,van_Ravenzwaaij_Cassey_Brown_2018,Shapiro_2003,Tokdar_Kass_2010,Baumert_Ghate_Kiatsupaibul_Shen_Smith_Zabinsky_2009}.
These techniques were used for sampling propositional\BCFV{and integer} formulas in~\cite{kitchen2007stimulus,kitchen2010markov}.
Additional algorithms for sampling propositional formulas are based on syntactic mutations~\cite{dutra2018efficient}, random walks~\cite{wei2004towards}, recursive search~\cite{ermon2012uniform}, knowledge compilation~\cite{sharma2018knowledge}, adaptation of SAT solvers and model counters ~\cite{Agbaria_Carmi_Cohen_Korchemny_Lifshits_Nadel_2010,nadel2011generating,achlioptas2018fast}, and universal hashing~\cite{meel2014sampling,meel2016constrained,ermon2013embed,Chakraborty_Meel_Vardi_2013}.

% Specifically related is the problem of uniform sampling of integer points in polyhedra~\cite{Meister_Clauss_2020,Pak_2002,Pouchet_Bastoul_Cohen_Vasilache_2007,Propp_Wilson_1998}
% is of interest in computational geometry as well as in enumerative combinatorics, algebraic geometry, and Applied Statistics
% In statistics, one often need to obtain many independent uniform samples of the integer
% points in certain polyhedra (e.g. the set of contingency tables) to approximate
% a certain distribution on them (e.g. $\xi^2$ distribution).
% \paragraph{Sampling of Polygons}
% Found this~\cite{cantarella2016fast}. \BC{TODO: sampling of polygons}
% They talk about sampling open/closed polygons. 
% Apparently there is a whole line of works on the subject. 
% Not sure if and how to relate to it.

% \paragraph{\textsc{SMTSampler}}
The problem of sampling for arbitrary SMT formulae was, however, much less explored.
A prominent exception is \SMTSampler~\cite{Dutra:2018}, which samples formulas at the SMT level.
Like \MEGASampler, \SMTSampler is epoch-based.
However, their way of extending a single model to a set of models relies on syntactic mutations and their combination.
% While these mutations may lead to alternative models of the formula, they are not guaranteed to do so.
% Hence, one has to check explicitly if the discovered assignments satisfy the formula or not.
% In contrast, \textsc{MegaSampler} relies on sampling a formula $\hat\varphi$ which is a model approximation of $\varphi$. 
% Models of this formula are guaranteed to be models of $\varphi$, by construction.
Another difference is that \SMTSampler was designed for formulas in the theory of bit-vectors with arrays and uninterpreted functions.
% Our framework, on the other hand, can be applied to any $T$ and $T'$ s.t. $T'$ restricts $T$.
% That being said, we believe that the core idea of \textsc{SMTSampler}, \ie using mutations to generate novel solutions, can be generalized to other theories as well.
In our experiments, we implemented a version of their algorithm for the $\Tmia$ and $\Tamia$ theories and compared it empirically against \MEGASampler.
A follow-up work on SMT sampling is \textsc{GuidedSampler}~\cite{Dutra:2019}, which is a variant of~\cite{Dutra:2018} that allows providing a problem-specific coverage metric and aims to optimize it.

% % \paragraph{\textsc{GuidedSampler}}
% A follow-up work on SMT sampling is \textsc{GuidedSampler}~\cite{Dutra:2019}.
% This work focuses on allowing the user to provide a problem-specific coverage metric.
% The underlying algorithm resembles that of \textsc{SMTSampler}: it is epoch-based and mutations are used to generate new models from the seed.
% Except, this variant is tailored towards maximizing the specified coverage criteria.
% While we agree that being able to specify a user-defined coverage criteria can be of value, this is not the focus of this paper.

% \paragraph{Related Problems in SMT}
% Related to model approximation:
% under-approximation of SMT formulas and its usage in software analysis.
% over-approximation for static analysis.

% Related problems in SMT? Model counting + approximation.
% Applications of Sampling? Fuzzing symbolic expressions

The theory of intervals, $\Tic$, has been used for approximation in abstract interpretation since the very beginning~\cite{cousot1977static}.
However, the use of intervals in that context is for over-approximation, while ours is for under-approximation.
Intervals have also been used in~\cite{huang2020pangolin} to improve the efficiency of solving bit-vector formulas, and in~\cite{choi2019grey}, to approximate path constraints for 
concolic testing.
The work of~\cite{borzacchiello2021fuzzing} uses fuzzing techniques on SMT formulas to increase their efficiency for fuzzing-related instances, and also makes use of intervals.
Other methods of approximation were used to improve SMT solving~\cite{yao2020fast,bryant2007deciding}.
However, to the best of our knowledge, this is the first work to use approximation for SMT sampling.
% Also, the use of a model to guide the approximation process is novel.

% Fuzzing symbolic expressions: In this paper, we investigate whether techniques borrowed from the fuzzing domain can be applied to check whether symbolic formu- las are satisfiable in the context of concolic and hybrid fuzzing engines, providing a viable alternative to classic SMT solving techniques.

The notion of Model-Based Projection (MBP)~\cite{Komuravelli_Gurfinkel_Chaki_2016} has some common ground with the notion of Model-Guided Approximation (MGA) (\cref{sec:mbua}), but there are also some significant differences between the two; see discussion in \cref{appendix:MBP}.

\BCFV{Todo: more references in comments: sampling of polyhedra}
% Discrete hit and run~\cite{Baumert_Ghate_Kiatsupaibul_Shen_Smith_Zabinsky_2009} - solve the problem of sampling a point distributed according to a general distribution $\pi$ over an arbitrary subset $S$ of an integer hyperrectangle $H$.
% They have 6 more refernces for this particular problem.
% In their work, both $S$ and $\pi$ are given as oracles (as opposed to explicit formulas or equations) which a. is useful for applications, b. makes the problem harder.

\section{Conclusion}\label{sec:conclusion}
We have shed some new light on the intriguing problem
of sampling from the set of satisfying assignments for
an SMT formula,
by offering an alternative to the existing stochastic
mutation-based approach.
The reduction to an intermediate theory, such as the
interval theories $\Tic,\Taic$ that we used in our
proof of concept and evaluation, sidesteps the need
for intensive generate-and-test cycles, as model-guided approximation is guaranteed to be an
underapproximation of the input formula from which we can freely sample without
having to check, and possibly discard, some of the
generated assignments.
In a sense, model-guided approximation ``squeezes the most'' out of each seed.
% , which is a very good thing, as seeds are hard to come by---they require MAX-SMT, a relatively expensive procedure.
Our evaluation shows that the new approach indeed
improves the performance of SMT sampling in practice.

\begin{comment}
The \(\texttt{MeGASample}^T_{T'}\) framework is fairly generic and
can be applied to more theories (we detail a few options in
\cref{appendix:additional-theories}).
Interval constraints lend themselves easily to sampling on account
of being non-relational, but with further mathematical development,
relational domains can be used to under-approximate, and then sample,
more complex formulas.
\end{comment}
\subsubsection*{Acknowledgements}
This work is supported by the Israeli Science Foundation Grant No. 243/19 and the Binational Science Foundation (NSF-BSF) Grant No. 2018675.

The authors would like to thank Profs. Orna Grumberg and Ofer Strichman for their valuable input and contributions to this work. We would additionally like to thank the anonymous reviewers for their time and effort.
\bibliography{main}
\clearpage
\appendix
\section{Handling Array Equality in $\Tamia$}
\label{appendix:array-eq}

In \cref{sec:arrays}, we introduced the theory $\Tamia$ of
formulas over arrays, with terms corresponding to selection
of elements in arrays and storing them.
We have shown that terms involving array sorts can always be reduced to a fragment that contains only $\aselect$, so long as
these terms are never at the top level, that is, are not direct
arguments to an atom's predicate.
Indeed, the only case where they \emph{can} be such, is when using the equality predicate over an array sort.

An equality atomic formula thus has the following general form:
\begin{equation}\label{eq:array-eq:form}
\begin{array}{l}
\astore(\astore(\cdots \astore(a_1, i_1, t_1), \cdots, i_{n-1}, t_{n-1}), i_n, t_n)
 = \\
 \qquad
\astore(\astore(\cdots \astore(a_2, j_1, s_1), \cdots, j_{m-1}, s_{m-1}), j_m, s_m)
\end{array}
\end{equation}

The implication of this is that $a_1$ and $a_2$ are equal everywhere except perhaps in (at most) $n + m$ locations whose indices are $i_1,\ldots,i_n,j_1,\ldots,j_m$.
Consequently, we can define a new array variable $c$ and argue that
\begin{equation}\label{eq:array-eq:subst}
\begin{array}{l}
a_1 = \astore(\cdots \astore(c, i_1, u_1), \cdots, i_n, u_n) \\
a_2 = \astore(\cdots \astore(c, j_1, v_1), \cdots, j_m, v_m)
\end{array}
\end{equation}

For $n + m$ new scalar-valued variables $u_1,\ldots,u_n,v_1,\ldots,v_m$, combined with the following
constraints:
\begin{equation}\label{eq:array-eq:selects}
\begin{array}{l}
  \aselect(c, i_k) = t_k \quad (k=1..n) \\
  \aselect(c, j_k) = s_k \quad (k=1..m)
\end{array}
\end{equation}

We can therefore transform the $m$-implicant $P$ by replacing the literal (\ref{eq:array-eq:form}), with the conjunction of (\ref{eq:array-eq:selects}), and substituting $a_1$ and $a_2$ elsewhere according to (\ref{eq:array-eq:subst}).

It only remains to handle inequalities of the form $a_1\neq a_2$; but these are never a problem, because arrays are infinite, so it is always possible to cause arrays to be unequal during sampling by selecting arbitrary values to some unconstrained elements in the participating arrays.

% \paragraph{The \texttt{ComputeImplicant} procedure}
\section{Computing an $m$-implicant of $\varphi$}\label{appendix:m-implicant}
Several algorithms exist for computing all prime implicants of $\varphi$ \cite{Coudert_Madre_Jaures_1993,Jackson_Pais_1990,Slagle_Chang_Lee_1970,Strzemecki_1992}.
However, we only need to find one implicant of $\varphi$ that need not be prime, but has to be satisfied by $m$.\BC{define prime just to say we don't care about it?}
Therefore, we follow this simple procedure:
Assuming the formula was a-priori converted to NNF, we traverse its abstract-syntax-tree in a top-down fashion, and from each disjunction we leave only one disjunct that is chosen randomly from all disjuncts satisfied by $m$.
The top-down order ensures that $m$ satisfies every sub-formula we encounter. 
This, in turn, ensures that for every disjunction we encounter there exists at least one disjunct satisfied by $m$.
It also ensures that every literal in the returned product term is satisfied by $m$.
\BC{Do we need the algorithm? proof? proof sketch? or is this enough?}
\BC{go back to the review and the Yices reference and see if a citation is needed}

\section{Detailed Measurements}
\label{appendix:results-full}

The table in \cref{sec:eval} (\cref{fig:mega-table}) contains summarized data where
sub-subfolders (specifically, of QF\_LIA/CAV2009-slacked, QF\_LIA/CAV2009, and QF\_NIA/VeryMax) were aggregated to make the amount of data more manageable.

This appendix contains more elaborated results shown in~\cref{fig:mega-table-full}.

\begin{table*}[t]
    \centering
    \begin{adjustbox}{width=\textwidth,center=\textwidth}
    \begin{threeparttable}
    \begin{tabular}{l | c c | r r r | r r r r | r r r}
        \toprule
        &  & & \multicolumn{3}{c|}{Coverage (\%)} & \multicolumn{3}{c}{Epochs} & \multicolumn{1}{c}{\#SMT} & \multicolumn{3}{|c}{Unique Solutions} \\
        Benchmarks & vars & depth & \MEGA & \MEGAb & \SMTint & \MEGA & \MEGAb & \SMTint & \SMTint & \MEGA & \MEGAb & \SMTint \\
        \midrule
        
        % benchmark a_ints b_depth c_coverage_MeGA c_coverage_MeGAb c_coverage_SMT d_epochs_MeGA d_epochs_MeGAb d_epochs_SMT e_smtcalls_SMT f_solutions_MeGA f_solutions_MeGAb f_solutions_SMT                                                                         
QF\_ALIA/qlock2                                    & {     484} & {     233} & \textbf {   87.90} & {   20.40} & {   26.25} & \textbf {       3} & {       2} & {       2} & {      23} & {   12332} & \textbf {   14877} & {      20} \\
QF\_LIA/CAV2009-slacked\tnote{1}\;\,/10-vars       & {      20} & {       5} & \textbf {   99.64} & {   68.52} & {   95.29} & {     344} & \textbf {    2054} & {    1520} & {   29591} & \textbf { 4239716} & { 3828207} & { 1969209} \\
QF\_LIA/CAV2009-slacked\tnote{1}\;\,/15-vars       & {      30} & {       5} & \textbf {   99.68} & {   65.76} & {   90.20} & {     313} & \textbf {    2101} & {     124} & {    3856} & \textbf { 2069132} & { 1928526} & { 1478326} \\
QF\_LIA/CAV2009-slacked\tnote{1}\;\,/20-vars       & {      40} & {       5} & \textbf {   98.67} & {   63.49} & {   84.27} & {     119} & \textbf {     240} & {      78} & {    3199} & { 1574733} & \textbf { 1985446} & {  969441} \\
QF\_LIA/CAV2009-slacked\tnote{1}\;\,/25-vars       & {      50} & {       5} & \textbf {   97.97} & {   68.31} & {   96.34} & {      88} & \textbf {     291} & {      50} & {    2527} & { 1247261} & \textbf { 1891225} & {  484204} \\
QF\_LIA/CAV2009-slacked\tnote{1}\;\,/30-vars       & {      60} & {       5} & \textbf {   98.41} & {   65.44} & {   97.69} & {      96} & \textbf {     173} & {      96} & {    5867} & { 1389891} & \textbf { 1685593} & {   77821} \\
QF\_LIA/CAV2009-slacked\tnote{1}\;\,/35-vars       & {      70} & {       5} & \textbf {   97.52} & {   64.26} & {   91.63} & {      66} & \textbf {     131} & {      36} & {    2556} & {  914320} & \textbf { 1520734} & {   57162} \\
QF\_LIA/CAV2009-slacked\tnote{1}\;\,/40-vars       & {      80} & {       5} & \textbf {   96.28} & {   65.88} & {   81.25} & {      60} & \textbf {      97} & {      29} & {    2334} & {  852702} & \textbf { 1320256} & {   24429} \\
QF\_LIA/CAV2009-slacked\tnote{1}\;\,/45-vars       & {      90} & {       5} & \textbf {   98.71} & {   67.21} & {   82.25} & {      60} & \textbf {      82} & {      46} & {    4038} & {  891446} & \textbf { 1167970} & {    3783} \\
QF\_LIA/CAV2009\tnote{3}\;\,/10-vars               & {      10} & {       5} & \textbf {   98.52} & {   87.27} & {   97.50} & {     448} & {    2255} & \textbf {    8328}
 & {   91606} & \textbf { 6432074} & { 5918818} & { 2254764} \\
QF\_LIA/CAV2009\tnote{3}\;\,/15-vars               & {      15} & {       5} & \textbf {   99.53} & {   92.09} & {   99.30} & {     465} & {    2171} & \textbf {    2335}
 & {   37353} & \textbf { 4370035} & { 4178125} & {  839561} \\
QF\_LIA/CAV2009\tnote{3}\;\,/20-vars               & {      20} & {       5} & \textbf {   98.51} & {   92.93} & {   98.44} & {     206} & {     658} & \textbf {     661}
 & {   13882} & { 2925180} & \textbf { 4308711} & {  459251} \\
QF\_LIA/CAV2009\tnote{3}\;\,/25-vars               & {      25} & {       5} & {   48.28} & \textbf {   98.90} & {   68.68} & {      74} & \textbf {     502} & {     115}
 & {    2977} & {  830669} & \textbf { 3783846} & { 1040005} \\
QF\_LIA/CAV2009\tnote{3}\;\,/30-vars               & {      30} & {       5} & {   43.98} & \textbf {   99.29} & {   66.42} & {      61} & \textbf {     676} & {      83}
 & {    2559} & {  657487} & \textbf { 3326873} & {  861596} \\
QF\_LIA/CAV2009\tnote{3}\;\,/35-vars               & {      35} & {       5} & {   44.73} & \textbf {   97.94} & {   65.51} & {      38} & \textbf {     468} & {      22}
 & {     797} & {  367092} & \textbf { 3044466} & { 1233888} \\
QF\_LIA/CAV2009\tnote{3}\;\,/40-vars               & {      40} & {       6} & {   57.54} & \textbf {   96.71} & {   73.54} & {      53} & \textbf {     562} & {      81}
 & {    3319} & {  582002} & \textbf { 2559974} & {  797306} \\
QF\_LIA/CAV2009\tnote{3}\;\,/45-vars               & {      45} & {       6} & {   35.25} & \textbf {   96.75} & {   63.01} & {      30} & \textbf {     329} & {      33}
 & {    1504} & {  280406} & \textbf { 2397593} & { 1034948} \\
QF\_LIA/bofill-sched-random\tnote{4}\;\,           & {     780} & {       6} & \textbf {   99.84} & {   83.95} & {   76.00} & {     628} & \textbf {     927} & {       4}
 & {     196} & {     627} & \textbf {     927} & {     112} \\
QF\_LIA/bofill-sched-real\tnote{5}\;\,             & {     576} & {       6} & \textbf {   99.94} & {   91.95} & {   81.74} & {     689} & \textbf {    1159} & {       3}
 & {     580} & {     688} & \textbf {    1158} & {     249} \\
 QF\_LIA/convert                                    & {     768} & {    1339} & {   30.61} & {   27.71} & \textbf {   80.24} & \textbf {       7} & {       6} & {       2}
 & {      65} & \textbf {   90635} & {   76671} & {      53} \\
QF\_LIA/dillig                                     & {      31} & {      25} & {   42.20} & \textbf {   97.70} & {   53.57} & {     133} & \textbf {    1164} & {     769}
 & {   10551} & { 1729099} & \textbf { 3983851} & {  455503} \\
QF\_LIA/pb2010                                     & {    5842} & {       5} & \textbf {   75.64} & {   71.85} & {   51.37} & {   30099} & {     107} & \textbf {   42907}
 & {  128875} & {      53} & \textbf {     214} & {      27} \\
QF\_LIA/prime-cone                                 & {      10} & {       5} & {   96.89} & {   63.86} & \textbf {   97.00} & {    2705} & {    9313} & \textbf {   39360}
 & {  155484} & \textbf {13658385} & { 3906854} & {  613784} \\
QF\_LIA/slacks                                     & {      61} & {       5} & \textbf {   99.11} & {   69.50} & {   85.75} & {     135} & {     205} & \textbf {     383}
 & {   11005} & { 1923319} & \textbf { 2245506} & {  335504} \\
QF\_NIA/20170427-VeryMax/CInteger                  & {     106} & {      11} & \textbf {   86.15} & {   57.44} & {   56.18} & {      15} & \textbf {     875} & {       3}
 & {      28} & {   65440} & \textbf { 1769769} & {    1479} \\
QF\_NIA/20170427-VeryMax/ITS                       & {     256} & {      15} & \textbf {   86.02} & {   51.49} & {   51.79} & {      11} & \textbf {     138} & {       2}
 & {      33} & {   98204} & \textbf { 1356607} & {   35918} \\
QF\_NIA/AProVE                                     & {      40} & {       6} & \textbf {   97.87} & {   71.57} & {   77.17} & {     714} & {    6016} & \textbf {   10053}
 & {   73190} & { 6478901} & \textbf { 6982253} & { 1394814} \\
QF\_NIA/leipzig                                    & {      92} & {       2} & {   59.55} & \textbf {   74.50} & {   54.22} & {      20} & \textbf {    1305} & {       3}
 & {      64} & {   96843} & \textbf { 1189523} & {    1207} \\
% coverage_MeGA=0.8149560974400868 solutions_MeGA=1884333.1151079137 coverage_MeGAb=0.7384711871197046 solutions_MeGAb=2272399.273381295 coverage_SMT=0.7706412352397715 solutions_SMT=559008.8393285371 coverage_megasampler=0.9421652649686911 solutions_megasampler=2078366.1942446043
        \bottomrule
    \end{tabular}
    \begin{tablenotes}
        \item[1] Short for \emph{CAV\_2009\_benchmarks/smt}
        \item[2] Short for \emph{20180326-Bromberger/more\_slacked/CAV\_2009\_benchmarks/smt}
        \item[3] Short for \emph{20180326-Bromberger/unbd-sage}
        \item[4] Short for \emph{bofill-scheduling/SMT\_random\_LIA}
        \item[5] Short for \emph{bofill-scheduling/SMT\_read\_LIA}
    \end{tablenotes}
    \end{threeparttable}
    \end{adjustbox}

    \vspace{5pt}
    \caption{Results (averaged) over the benchmarks}
    \vspace{-5pt}
    \label{fig:mega-table-full}
\end{table*}

\section{Model-Based Projection}\label{appendix:MBP}
The work on Model-Based Projection (MBP)~\cite{Komuravelli_Gurfinkel_Chaki_2016} is related to the notion of Model-Guided Approximation, defined in~\cref{sec:mbua}.
Given a formula $\eta(y)=\exists \eta_m(x,y)$, Model-Based Projection (MBP) is defined as a function $Proj_\eta$ from models of $\eta_m$ to quantifier-free formulas over $y$, with the following requirements:
\begin{enumerate}
    \item $Proj_\eta$ has a finite image
    \item $\eta \equiv \bigvee_{M \models \eta_m} Proj_\eta(M)$
    \item for every model $M$ of $\eta_m$, $M \models Proj_\eta(M)$
\end{enumerate}

Behind this definition and the definition of an $M$-approximation (\cref{def:m-approximation}) is a similar idea, of using a model to guide the process of under-approximation of a formula for a particular purpose.
In MBP, $Proj_\eta(M)$ under-approximates $\eta$, and in MGA, an $M$-approximation $\varphi'$ under-approximates $\varphi$.
However, the purpose is quite distinct.
In MBP, the restriction on $Proj_\eta(M)$ stems from the fact that it should only be over $y$, and from the fact that the image of $Proj_\eta$ should be finite and should cover $\eta$ (requirements 1-2). 
In MGA, however, the restriction on $\varphi'$ stems from the requirement that $\varphi'$ be in $T'$.
% In particular, for MBP, a trivial solution will always be to set  to the formula itself, while mapping each model to itself will not necessarily work.
% On the other hand, for MGA, the situation is reversed: a formula encoding the model itself will always be a trivial solution (assuming theory T 'allows for equations),
% While the formula itself will usually not constitute a solution (will not belong to T ').

An interesting future work direction is thus to generalize the idea of using a model to guide an under-approximation process for other purposes as well.

\section{Proof of the Grounding Fidelity Lemma (\Cref{lemma:grounding_fidelity})}\label{appendix:grounding-proof}
\begin{proof}
By induction on the structure of terms $t\sqsubseteq\varphi$.
If $t$ is a select-term, then $t\in S$ and $m[t]=\agm[v_t]$ by the definition of atomic grounding.
Otherwise, $t$ is of the form $f(t_1,\ldots,t_n)$ where $f$ is an interpreted function of the theory and $t$ is not of array sort.
(Notice that subterms in $\aselect$ are never reached in this proof, and all other occurrences of arrays have been eliminated.)
Therefore the equivalence is obtained from the induction hypothesis on $t_1,\dots,t_n$ and from the fact that the interpretation of $f$ is fixed.
\end{proof}
%\section{Proofs}
%\input{proofs}

\iffunky
\setcounter{tocdepth}{1}
\listoftodos
\fi

\end{document}